\documentclass{aa}  

\usepackage{graphicx}
\usepackage{pstricks}
\usepackage{color}

\usepackage{amsmath}
\DeclareMathOperator{\asin}{asin}

\usepackage{wasysym}

\usepackage{txfonts}
\usepackage{hyperref}
\hypersetup{
    bookmarks=true,         
    unicode=true,           
    pdftoolbar=true,        
    pdfmenubar=true,        
    pdffitwindow=true,      
    pdftitle={Solar System Science with ESA Euclid},
    pdfauthor={B. Carry},
    pdfsubject={Planetary Science},
    pdfkeywords={},         
    pdfnewwindow=true,      
    colorlinks=true,        
    linkcolor=gray,         
    citecolor=blue,         
    filecolor=gray,         
    urlcolor=gray           
}

\usepackage{enumitem}
\usepackage{ulem}

\graphicspath{{figures/}}

\newcommand{\now}{\numb{2017 June 28}}
\newcommand{\tsub}[1]{$_{\textrm{#1}}$}
\newcommand{\tup}[1]{$^{\textrm{#1}}$}

\newcommand{\numb}[1]{#1}
\newcommand{\add}[1]{#1}

\newcommand{\nObs}{150,000}

\begin{document}

  \title{Solar System Science with ESA Euclid}

  \author{B. Carry\inst{1,2}
  }

  \institute{%
    Universit\'e C{\^o}te d'Azur, Observatoire de la C{\^o}te d'Azur, CNRS, Lagrange, France
    \and
    IMCCE, Observatoire de Paris, PSL Research University, CNRS,
    Sorbonne Universit{\'e}s, UPMC Univ Paris 06, Univ. Lille, France
  }

   \date{Received September 15, 1996; accepted March 16, 1997}

 
  \abstract
   {
     The ESA Euclid mission \add{has been designed to map the
       geometry of the dark Universe. Scheduled for launch in 2020, it}
     will conduct \add{a six-years} visible and 
     near-infrared imaging and spectroscopic survey over \numb{15,000}
     deg$^2$ down to V\tsub{AB}$\sim$\numb{24.5}.
     Although the survey will avoid ecliptic latitudes below
     \numb{15}\degr, the survey pattern in repeated sequences of
     four broad-band filters seems well-adapted to Solar
     System objects (SSOs) detection and characterization.
   }
   {
     We aim at evaluating Euclid capability to discover SSOs, and
     measure their position, apparent magnitude, and spectral energy distribution.
     Also, we investigate how these measurements can lead to the
     determination of their orbits,
     \add{morphology (activity and multiplicity), }
     physical properties (rotation period, spin
     orientation, \add{and} 3-D shape), and surface composition.
   }
   {
     We use current census of SSOs to extrapolate the total amount of
     SSOs detectable by Euclid, i.e., within the survey area and
     brighter than the limiting magnitude.
     For each different population of SSO, from neighboring near-Earth
     asteroids to
     distant Kuiper-belt objects (KBOs) and including comets, we compare the
     expected Euclid astrometry, photometry, and spectroscopy with SSO
     properties to estimate
     how Euclid will constrain the SSOs dynamical, physical, and
     compositional properties.
   }
   {
     With current survey design, \add{about 150,000} SSOs, mainly from the
     asteroid main-belt, should be
     observed by Euclid. These objects will all
     have high inclination, which contrasts with many SSO surveys
     focusing on the ecliptic plane.
     There is a potential for discovery of several 10$^4$ SSOs
     by Euclid, \add{in particular distant KBOs at high declination.}
     Euclid observations, consisting in a suite of four sequences of
     four measurements, will refine the spectral
     classification of SSOs by extending the spectral coverage
     provided by, e.g., Gaia
     and the LSST to 2 microns.
     The time-resolved photometry, combined with sparse photometry such as
     measured by Gaia and the LSST, will contribute to the
     determination of SSO rotation period, spin orientation,
     and 3-D shape model.
     The sharp and stable point-spread function of Euclid will also
     allow to resolve binary systems in the Kuiper Belt and detect
     activity around Centaurs.
   }
   {
     The depth of Euclid survey (V\tsub{AB}$\sim$\numb{24.5}), 
     its spectral coverage (0.5 to 2.0\,$\mu$m), 
     and observation cadence has great potential for Solar System
     research.
     A dedicated processing for SSOs is being set in place
     within Euclid consortium to produce 
     catalogs of astrometry, multi-color and time-resolved photometry, and
     spectral classification of some 10$^5$ SSOs, delivered as Legacy
     Science. 
   }


   \maketitle
%


\section{Introduction}

\begin{figure*}[ht]
  \centering
  \includegraphics[width=0.8\hsize]{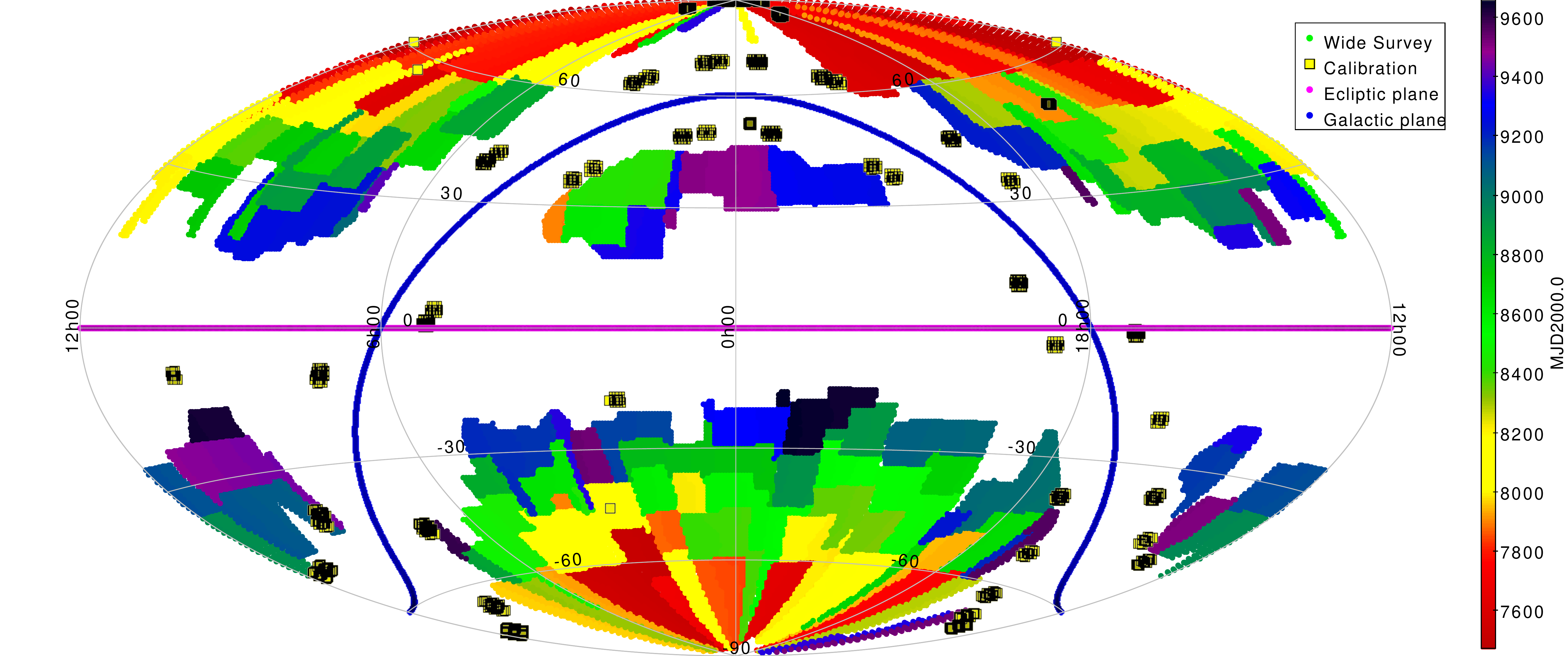}
  \caption[Euclid Reference Survey]{
    Expected coverage of the Euclid \textsl{Wide survey}
    (called the reference survey), color-coded by observing epoch,
    in an Aitoff projection of ecliptic coordinates.
    The horizontal gap corresponds to low ecliptic latitudes
    \add{(the cyan line represents the ecliptic plane)}, and the
    circular gap to low galactic latitudes \add{(the deep blue line
      stands for the galactic plane)}.
    The \add{black squares filled with yellow} are the calibration
    fields, repeatedly observed 
    over the six years of the mission,
    \add{to assess the stability and accuracy of Euclid point-spread
      function (PSF), photometry, and spectroscopy.}
  }
  \label{fig:rs}
\end{figure*}

\begin{figure}[ht]
  \centering
  \includegraphics[width=0.8\hsize]{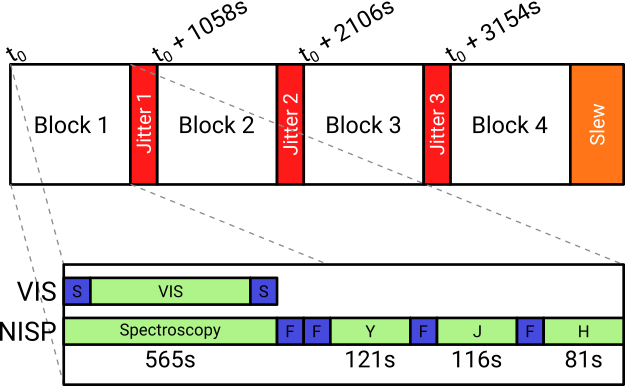}
  \caption{
    Observation sequence for each pointing.
    The observing block composed by a simultaneous VIS and
    NISP/spectroscopy exposure, and three NISP/imaging exposures (Y,
    J, H) is repeated four times, with small jitters
      (\numb{100\arcsec\,$\times$\,50\arcsec}).
    Blue boxes F and S stand for overheads due to the rotation of the
    filter wheel and shutter opening/closure.
    Figure adapted from \citet{2011-Euclid-Laureijs}
  }
  \label{fig:seq}
\end{figure}

  \indent Euclid, the second mission in ESA's Cosmic Vision program, 
  is a wide-field space mission dedicated to the
  study of dark energy and dark matter through a mapping of weak
  gravitational lensing \citep{2011-Euclid-Laureijs}. 
  It is equipped with a silicon-carbide 1.2\,m-aperture Korsch
  telescope and two instruments: 
  a VISible imaging camera and a Near Infrared 
  Spectrometer and Photometer \citep[VIS and NISP,
    see][]{2014-SPIE-Cropper, 2014-SPIE-Maciaszek}.
  The mission design combines a large field of view (FoV, \numb{0.57}
  deg$^2$) with high angular resolution
  (pixel scales of \numb{0.1}\arcsec~and \numb{0.3}\arcsec~for VIS and
  NISP, corresponding to the diffraction limit at 0.6 and 1.7\,$\mu$m).\\
  \indent Scheduled for a launch in 2020 and operating during six years
  from the Sun-Earth Lagrange L2 point, Euclid will
  carry out an imaging and spectroscopic survey of the extra-galactic
  sky of \numb{15,000} deg$^2$ (the \textsl{Wide Survey}),
  avoiding galactic latitudes smaller than
  \numb{30}\degr~and ecliptic latitudes below
  \numb{15}\degr~(Fig.~\ref{fig:rs}), totaling \numb{35,000} pointings.
  A second survey, two magnitudes deeper and located at very high
  ecliptic latitudes, will cover \numb{40} deg$^2$ spread in three areas
  (the \textsl{Deep Survey}). 
  Additionally, \numb{7,000} observations of \numb{1,200} calibration
  fields, 
  mainly located at -10\degr~and +10\degr~of galactic latitude,
  will be acquired over the course of the
  mission to monitor the stability of the telescope point-spread
  function (PSF), and assess the mission photometric and
  spectroscopic accuracy.\\
%
%
  \indent Euclid imaging detection limits are required at m\tsub{AB}\,=\,24.5
  (10\,$\sigma$ on a 1\arcsec~extended source) with VIS, and
  m\tsub{AB}\,=\,24 (5\,$\sigma$ point source) in the Y, J, and H
  filters with NISP.
  Spectroscopic requirements are to cover the same near-infrared
  wavelength range
  at a resolving power of \numb{380}
  and to detect at 3.5\,$\sigma$ an
  emission line at $3.10^{-16}$\,erg.cm$^{-1}$.s$^{-1}$
  (on a 1\arcsec~extended source).
  The NISP implementation consists in two grisms,
  \textsl{red} (1.25 to 1.85\,$\mu$m) and \textsl{blue} (0.92 
  to 1.25 $\mu$m, which usage will be limited to the \textsl{Deep
    Survey}), providing a 
  continuum sensitivity to m\tsub{AB}\,$\approx$\,\numb{21}.
  To achieve these goals, the following survey operations were
  designed:
  \begin{enumerate}
    \item The observations will consist in a step-and-stare tiling 
      mode, in which both instruments target the common 0.57 deg$^2$
      field of view before the telescope slews to
      other coordinates.
    \item Each tile will be visited only once, with the exception of the
      \textsl{Deep Survey} in which each tile will be
      pointed \numb{40} times, and the 
      calibration fields, observed \numb{5} times each on average.
    \item The filling pattern of the survey will follow lines of
      ecliptic longitude at quadrature.
      Current survey planning foresees a narrow distribution of solar 
      elongation of $\Psi$\,=\,91.0\,$\pm$\,1.5\degr~only, the range of solar
      elongation available to the telescope being
      limited to \numb{87}\degr--\numb{110}\degr.
    \item The observation of each tile will be sub-divided in
      four observing blocks, differing only by small jitters
      (\numb{100\arcsec\,$\times$\,50\arcsec}). These small pointing
      offsets will allow to fill the gaps between the detectors
      composing each instrument focal plane, resulting in 95\% of the
      sky covered by three blocks, and 50\% by four.
    \item In each block, near-infrared slitless spectra 
      will be obtained 
      with NISP simultaneously to a visible image with VIS,
      with an integration time of \numb{565}\,s.  
      This integration time implies a saturation limit of
      V\tsub{AB}$\approx$\numb{17} for a point-like source.
      Then, three NISP images will be taken with the Y, J, and H
      near-infrared filters, with integration time of
      \numb{121}, \numb{116}, and \numb{81}\,s respectively
      (Fig.~\ref{fig:seq}) .
  \end{enumerate}
%
%
%
%
  \indent All these characteristics make Euclid a 
  potential prime data set for legacy science.
  In particular, the access to the near-infrared sky,
  about \numb{7} magnitudes fainter than
  DENIS and 2MASS
  \citep{1994-ApSS-217-Epchtein, 2006-AJ-131-Skrutskie} surveys,
  and \numb{2--3} magnitudes fainter than current ESO VISTA
  Hemispherical Survey 
  \citep[VHS,][]{2013-Messenger-154-McMahon}, 
  makes Euclid appealing for surface characterization of 
  Solar System Objects (SSOs),
  especially in an era rich in surveys operating in visible
  wavelengths only such as 
  the Sloan Digital Sky Survey (SDSS), 
  Pan-STARRS, ESA Gaia, and the Large Synoptic Sky Survey (LSST)
  \citep{2003-AJ-126-Abazajian,2003-EMP-92-Jewitt,
    2016-AA-595-Prusti, 2009-Book-LSST}.\\
  \indent We discuss here the potential of the Euclid mission for Solar
  System Science. In the following, we consider the following populations of
  SSOs, defined by their orbital elements (Appendix~\ref{app:class}): 
  \begin{itemize}
    \item[$\circ$] the near-Earth asteroids (NEAs), including the
      Aten, Apollo, and Amor classes,
      which orbits cross that of terrestrial planets;
    \item[$\circ$] the Mars-crossers (MCs), a transitory population
      between the asteroid main belt and the near-Earth space;
    \item[$\circ$] the main-belt asteroids (MBA), in the principal
      reservoir of asteroids in the Solar System, between Mars and
      Jupiter, split into
      Hungarian, Inner Main-Belt (IMB), Middle Main-Belt (MMB), Outer
      Main-Belt (OMB), Cybele, and Hilda;
    \item[$\circ$] the Jupiter trojans (Trojans), orbiting the Sun at
      the Lagrange L4 and L5 points of the Sun-Jupiter system;
    \item[$\circ$] the Centaurs, which orbits cross that of giant
      planets; 
    \item[$\circ$] the Kuiper-belt objects (KBOs), further than
      Neptune, divided into Detached, Resonant, and Scattered-Disk
      Objects (SDO), and Inner, Main, and Outer Classical Belt (ICB,
      MCB, OCB); and
    \item[$\circ$] the comets, from the outskirts of the
      solar system, on highly eccentric orbits, and characterized by
      activity (presence of coma) at short heliocentric distances.
  \end{itemize}

  \indent The discussion is organized as following: 
  the expected number of observation of Solar System Objects by Euclid
  is presented in Section~\ref{sec:disco}, 
  and their challenges in Section~\ref{sec:detect}.
  The issue of source identification, and contribution to astrometry
  and orbit determination is discussed in Section~\ref{sec:astro}.
  Then the potential for spectral characterization from VIS and NISP
  photometry is detailed in Section~\ref{sec:taxo}, and from NISP
  spectroscopy in Section~\ref{sec:spec}.
  Euclid capabilities to directly image satellites and activity of
  SSOs are presented in Section~\ref{sec:psf}, and its contribution to
  3-D shape and binarity modeling from lightcurves in
  Section~\ref{sec:photo}.

\section{Expected number of SSO observations\label{sec:disco}}
\begin{figure*}
  \centering
  \includegraphics[width=0.49\hsize]{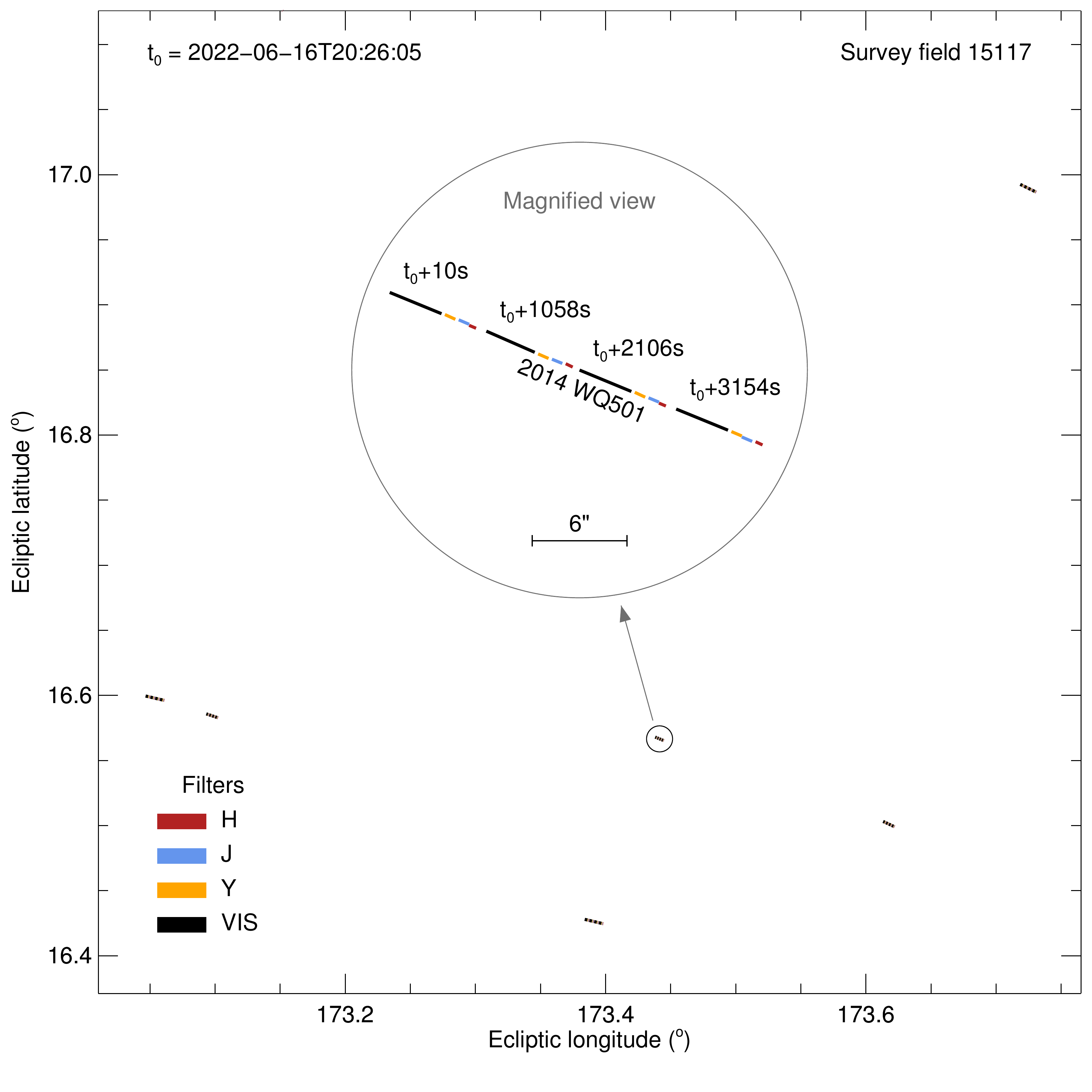}%
  \includegraphics[width=0.49\hsize]{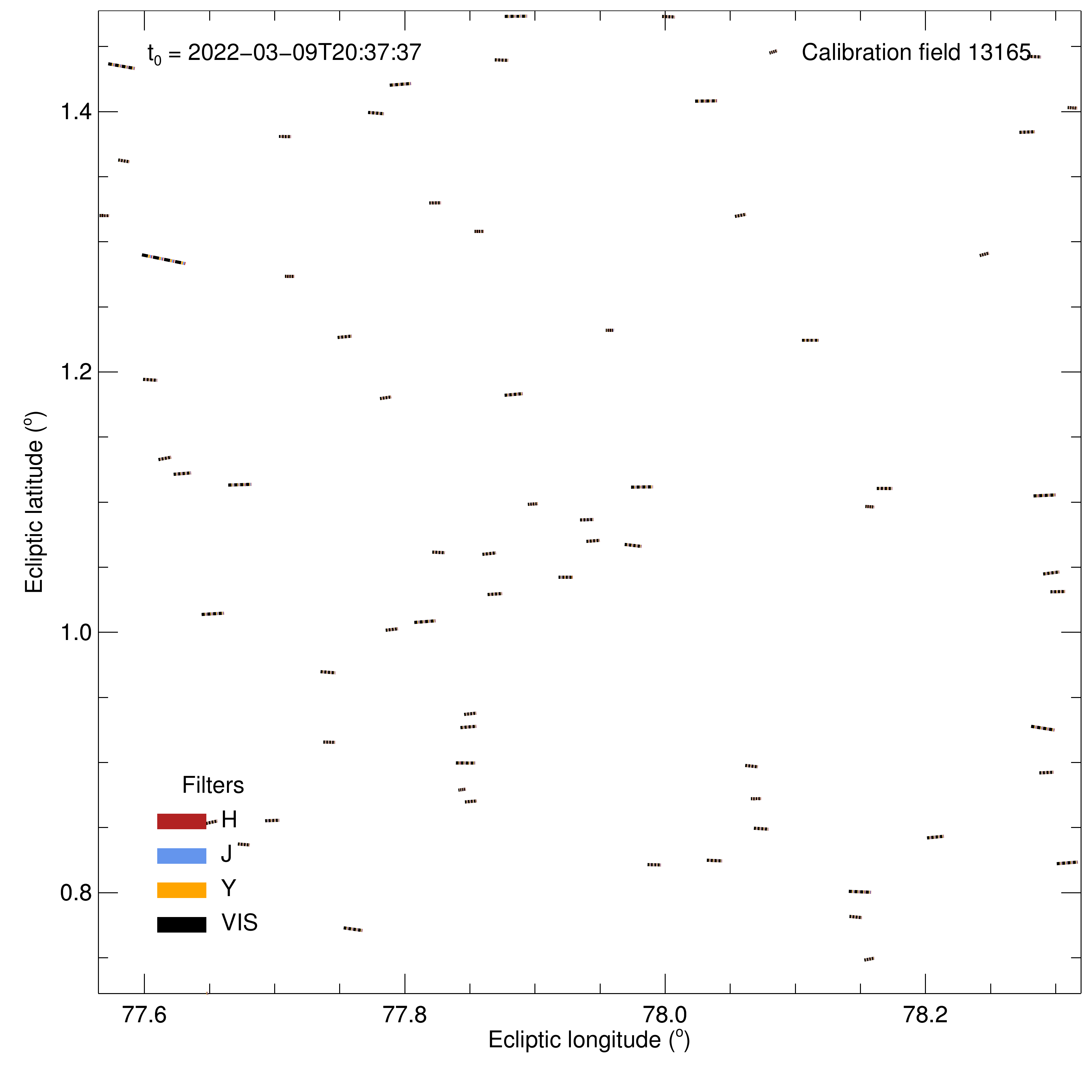}
  \caption[Example of SSO motion on Euclid stacked images]{
    Examples of the contamination of Euclid field of view by SSOs.
    \textbf{Left:} 
    Survey field \#15117 centered on
    (RA,Dec)\,=\,(167.218\degr,+12.740\degr) and starting 
    on 2022, June the 16\tup{th} at 20:26:05 UTC.
    The successive trails impressed
    by the 6 known SSOs during the Euclid hour-long sequence of
    VIS-NISP imaging 
    observations are drawn in different colors, one for each filter
    (VIS, Y, J, and H).
    We can expect about a hundred times more SSOs at Euclid limiting
    magnitude (e.g., Fig.~\ref{fig:sfd}).
    The inset is a magnified view of 2014 WQ501, a main-belt asteroid,
    illustrating the highly elongated shape of SSO in Euclid frames.
    The scale bar of 6\arcsec~corresponds to 60 pixels in VIS frames
    and 20 pixels in NISP.
    The timings reported are the starting time of VIS exposures.
    The slitless spectra will be acquired by NISP simultaneously to
    the VIS images.
    \textbf{Right:}
    Calibration field \#13165 centered on
    (RA,Dec)\,=\,(76.785\degr,+23.988\degr) and starting
    on 2022, March the 9\tup{th} at 20:37:37 UTC. There are 117 known
    SSOs in the field, and 
    here also, a hundred times more SSOs shall be detected at Euclid
    limiting magnitude.
  }
  \label{fig:img}
\end{figure*}

\begin{figure}[t]
  \centering
  \includegraphics[width=\hsize]{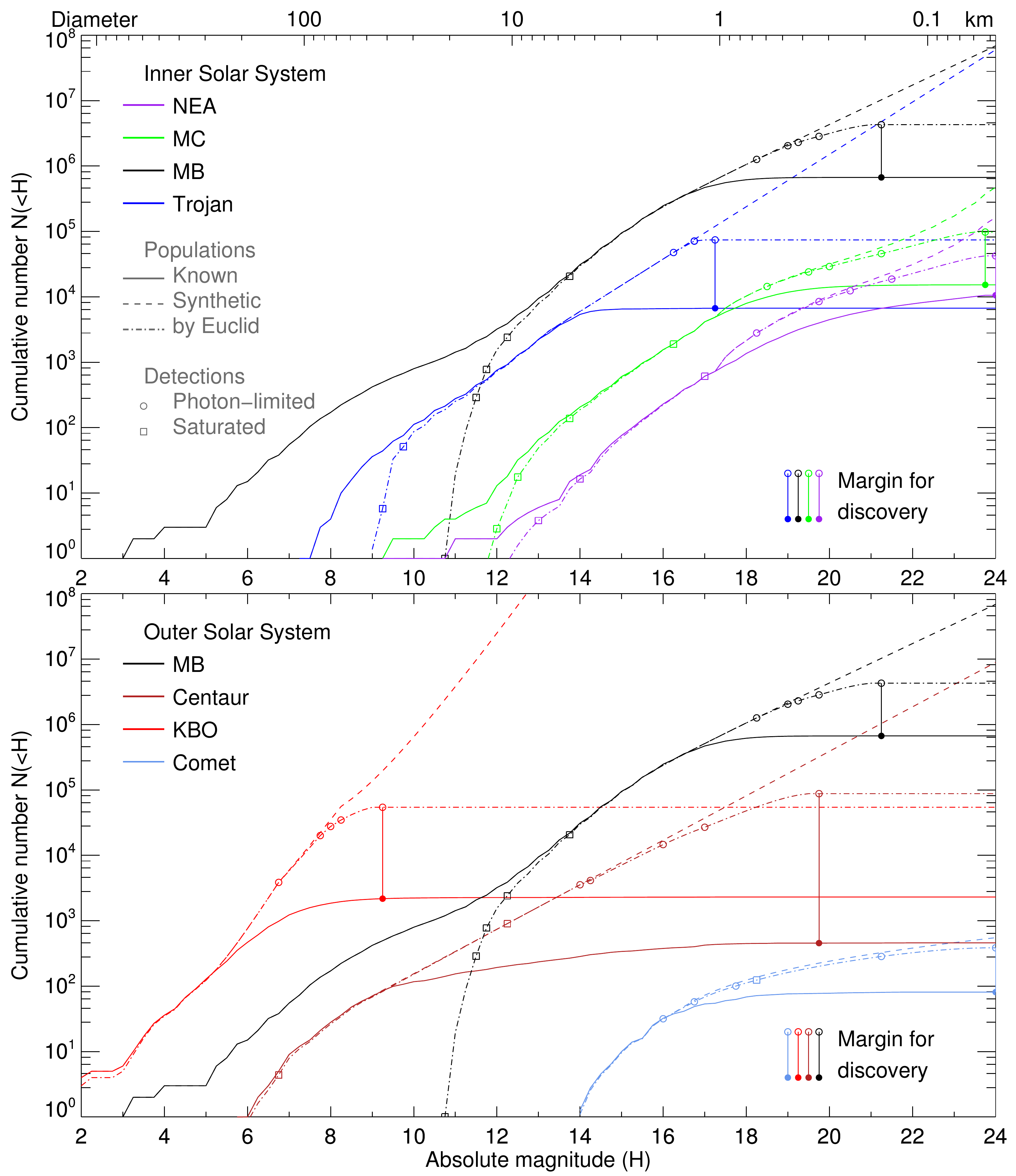}
  \caption[Cumulative size distribution]{
    Cumulative size distribution of each SSO population, for current
    census (solid lines) and synthetic populations (average estimates
    represented by the dashed lines, upper and lower estimate not
    plotted for clarity).
    The number of known objects, observable at Euclid limiting apparent
    magnitude over the entire celestial sphere, are
    represented by the dot-dashed lines.
    The open squares and circles represent the 0-25-50-75-100\% marks
    of the (H-V) cumulative probability function at the saturation and
    photon-starving ends.
    The total number of objects expected on the sky are marked by
    the filled circles. The difference between these filled
    circles and the current census represents the margin for discovery.
  }
  \label{fig:sfd}
\end{figure}

  \indent Although Euclid \textsl{Wide} survey will avoid the ecliptic plane
  (Fig.~\ref{fig:rs}),
  its design is casually very much adapted to detect moving objects. 
  As described above, each FoV will be imaged 16 times in one hour, in
  four repeated blocks.
  Given the pixel scale of VIS and NISP cameras of 0.1\arcsec~and
  0.3\arcsec, 
  any SSO with an apparent motion larger than $\approx$0.2\arcsec/h
  should
  therefore be detected by its trailed appearance and/or motion
  across the different frames (Fig.~\ref{fig:img}). \\
%
%
%
  \indent To estimate the number of SSOs that could be detected by
  Euclid, we first build the cumulative size distribution (CSD) of
  each population.
  We use the absolute magnitude H as a proxy for the diameter $D$. 
  The relation between both being 
  $D(\textrm{km}) = 1329 p_V^{-1/2} 10^{-0.2 H}$
  \citep[e.g.,][]{1989-AsteroidsII-Bowell}, where $p_V$ is the albedo
    of the surface in V, which quantifies its capability to reflect light.
  Minor planets, especially asteroids, tend to be very dark, and their
  albedo is generally very low, from a few percents to $\approx$30\%
  \citep[see, e.g.,][]{2011-ApJ-741-Mainzer}.  \\
  \indent We retrieve the absolute magnitude from the \texttt{astorb}
  database \citep{1993-LPI-Bowell}, with the exception of comets, not listed
  in \texttt{astorb}, for which we use the compiled data by
  \citet{2011-MNRAS-414-Snodgrass}.
  The challenge is then to extrapolate the observed distributions
  (shown as solid lines in Fig.~\ref{fig:sfd}) to smaller
  sizes. Most are close to power-law distributions
  \citep{1969-JGR-74-Dohnanyi} in the form
  $dN/dH \propto 10^{\gamma H}$, with different \textsl{slope} $\gamma$. 
  In the following, we model each population as following, and
  represent them with dashed lines in Fig.~\ref{fig:sfd}:
  \begin{itemize}
    \item[$\bullet$] NEAs:
      We use the synthetic population by
      \citet{2016-Nature-530-Granvik} which is very similar to the one
      by \citet{2015-Icarus-257-Harris}. 
      We, however, take a conservative approach and increase the
      uncertainty of the model to encompass 
      both estimates.
    \item[$\bullet$] MCs: There is no dedicated study of the CSD of
      MCs in the literature. We thus take the NEA model above, scaled
      by a factor of three to match currently known MC population.
      The upper estimate is taken as a power-law fit to current
      population with $\gamma$=\numb{0.41},
      and the lower estimate is the scaled NEA
      model by \citet{2016-Nature-530-Granvik}, reduced by a factor of two.
    \item[$\bullet$] MBAs: We use the \textsl{knee} distribution by
      \citet{2009-Icarus-202-Gladman}, in which large objects (H $\in$
      [11,15]) follow a steep slope ($\gamma \sim 0.5$) while
      smaller asteroids follow a shallower slope of $\gamma = 0.30 \pm 0.02$
      in the  range H $\in$ [15, 18], after which no constrain is
      available. 
      This model is scaled to \numb{25\,954} asteroids at H\,=\,\numb{15}.
      These authors found the CSD to be very smooth in that absolute
      magnitude range, compared to earlier works
      \citep{1998-Icarus-131-Jedicke, 2001-AJ-122-Ivezic,
        2007-AJ-133-Wiegert}.
      We only slightly modify their model, changing the slope 
      at H=15.25 instead of H=15: the shallower slope does not fit
      the observed data below H=15.25 anymore. The observing strategy by 
      \citet{2009-Icarus-202-Gladman} was indeed aimed at constraining the
      faint end of the CSD, and the constraints on large bodies
      was weak (only a small sky area had been targeted).
    \item[$\bullet$] Trojanss: We use the model of
      \citet{2000-AJ-120-Jewitt}, with $\gamma = 0.4 \pm 0.06$. More
      recently, \citet{2011-ApJ-742-Grav} found a similar $\gamma$,
      but restricted their study to Trojans with $D>10$\,km.
      We scale their model to the number of \numb{310} known Trojans at
      H\,=\,\numb{12.5}. 
      The steeper slope (i.e., $\gamma = 0.46$) seems to reproduce
      more accurately current observed population. The baseline numbers for Trojan
      presented here may therefore be underestimated, and the upper
      estimate could represent better the real Trojan population.
      Finally, we do not use the knee model by \citet{2005-AJ-130-Yoshida}, who
      predicted a change of slope at D$\approx$5\,km, because their
      model does not fit the known population anymore.
%
    \item[$\bullet$] Centaurs: We use the $\gamma = 0.34 \pm 0.04$
      from \citet{2013-ApJ-773-Bauer}, which is close from the 
      0.4 estimate from \citet{2002-AsteroidsIII-Jedicke}.
      We scale the power-law to correspond to the cumulative
      population (\numb{7}) at H\,=\,\numb{8.25}.
    \item[$\bullet$] KBOs:
      First, we build the CSD of the Resonant population using a
      single power-law of index $\gamma = 0.9_{-0.4}^{+0.2}$, scaled
      to a total of \numb{22,000} objects a H\,=\,\numb{8.66}, proposed by
      \citet{2016-AJ-152-Volk} based on the early results of the
      Outer Solar System Origins Survey
      \citep[OSSOS,][]{2016-AJ-152-Bannister}
      which is consistent with the earlier work by
      \citet{2012-AJ-144-Gladman}
      based on the Canada-France Ecliptic Plane Survey (CFEPS).
      Then, we build the CSD of the Scattered-disk objects
      using the \textsl{divot} distribution by
      \citet{2016-AJ-151-Shankman}:
      large objects follow a steep slope ($\gamma \sim 0.9$), scaled
      to a total of \numb{6500} objects a H\,=\,\numb{8}, which
      changes at H\,=\,8.0 to a shallower
      $\gamma = 0.50_{-0.08}^{+0.15}$. The differential size distribution 
      present a drop at H\,=\,8.0 where the slope changes, the smaller 
      objects being less numerous by a factor of \numb{5.6}
      \citep[see][for details]{2016-AJ-151-Shankman}.
      Finally, we take the CSD of objects in the Classical Belt from 
      \citet{2016-DPS-Petit} which propose a \textsl{knee}
      distribution:
      $\gamma=1.02$, scaled
      to a total of \numb{1800} objects at H\,=\,\numb{7}, until H=7.0 
      \citep[in agreement with][]{2014-AJ-148-Adams}
      where it switches to $\gamma=0.65 \pm 0.05$.
      The CSD for the entire KBO population is the sum of the three
      aforementioned CSD.
    \item[$\bullet$] Comets: We use the \textsl{knee} CSD from
      \citet{2011-MNRAS-414-Snodgrass}.
      Largest comets follow an $\gamma = 0.38_{-0.04}^{+0.06}$ until
      H\,=\,17 (converted from the turnover radius of 1.25 km using an
      albedo of 0.04) after which the CSD is shallower, although less
      constrained, and we assume the average slope found by
      \citet{2011-MNRAS-414-Snodgrass} with arbitrary uncertainties:
      $\gamma = 0.04_{-0.02}^{+0.06}$.
  \end{itemize}
  
  \indent The question is then what range of absolute magnitude will
  be accessible to Euclid for each population, considering it
  will observe in the range V\tsub{AB}\,=\,\numb{17--24.5}.
  This conversion from apparent to absolute magnitude
  only depends on the geometry of observation
  \citep{1989-AsteroidsII-Bowell} through the
  heliocentric distance ($\Delta$), range to observer ($r$), and phase
  angle ($\alpha$, the angle between the target-Sun 
  and target-observer vectors):

  \begin{equation}
    H = V + 2.5 \log\Big(r^2\Delta^2\Big) - 2.5 \log\Big( (1-G) \phi_1 + G\phi_2 \Big)
    \label{eq:H}
  \end{equation}

  with the phase functions approximated by
  \begin{eqnarray}
    \phi_1 &=& \exp\left(\ -3.33 \tan\left(\frac{\alpha}{2}\right)^{0.63} \right) \\
    \phi_2 &=& \exp\left(\ -1.87 \tan\left(\frac{\alpha}{2}\right)^{1.22} \right)
  \end{eqnarray}

  \indent Although a more accurate model (with two phase slopes $G_1$
  and $G_2$) of the phase dependence has been 
  developed recently \citep{2010-Icarus-209-Muinonen}, the differences in 
  the predicted magnitudes between the two systems are minor for our
  purpose.  
  We thus use the former and simplier H-G system in the following, assuming the
  canonic value of $G =\numb{0.15}$.\\
  \indent The three geometric parameters ($r$,$\Delta$,$\alpha$) are
  tight together by the solar elongation $\Psi$, which is imposed by the
  spacecraft operations ($\Psi$\,=\,91.0\,$\pm$\,1.5\degr).
  In practice, it is sufficient to
  estimate the range of heliocentric distances at which Euclid will
  observe an SSO from a given population to derive the two other
  geometric quantities, and hence the (H-V) index: 
  \begin{eqnarray}
    r      &=& \cos{\Psi} + \sqrt{ \cos^2{\Psi} -1 + \Delta^2. } \label{eq:r}\\
    \alpha &=& \left| \asin\left( \frac{\sin\Psi}{\Delta} \right)
    \right| \label{eq:phase}
  \end{eqnarray}

\begin{table*}[!t]
  \caption[Expected number of SSOs observed by Euclid]{
    Expected number of SSOs observed by Euclid for each population.
    For the whole celestial sphere, we report the current number of
    known SSOs ($\mathcal{N}$\tsub{now}, at the time of the
    writing on \now), the expected
    number of objects observable ($\mathcal{N}$\tsub{S})
    at Euclid limiting apparent magnitude
    (V\tsub{AB}$<$\numb{24.5}) and solar elongation
    ($\Psi$\,=\,91.0\,$\pm$\,1.5\degr). 
    Using the fraction of known SSOs present within the
    area of the Euclid \textsl{Wide} survey ($f_W$) and calibration frames
    ($f_C$), we estimate total number of 
    discoveries ($\mathcal{N}$\tsub{E,d}) and 
    observations ($\mathcal{N}$\tsub{E,o}) by Euclid.
    The absolute magnitude corresponding to a probability of
    100\%, 50\%, and 1\% that SSOs will be within Euclid detection
    envelop are also reported.
  }
  \label{tab:num}
  \centering
  \begin{tabular}{lrrrrrrrrr}
    \hline\hline
    \multicolumn{2}{c}{Population} & 
    \multicolumn{1}{c}{All-Sky} &
    \multicolumn{1}{c}{$f_W$} &
    \multicolumn{1}{c}{$f_C$} &
    \multicolumn{2}{c}{Euclid} &
    \multicolumn{3}{c}{Absolute magnitude limits} \\
    Name &
    \multicolumn{1}{c}{$\mathcal{N}$\tsub{now}} &
    \multicolumn{1}{c}{$\mathcal{N}$\tsub{S}} &
    \multicolumn{1}{c}{(\%)} &
    \multicolumn{1}{c}{(\%)} &
    \multicolumn{1}{c}{$\mathcal{N}$\tsub{E,d}} &
    \multicolumn{1}{c}{$\mathcal{N}$\tsub{E,o}} &
    \multicolumn{1}{c}{H\tsub{100}} &
    \multicolumn{1}{c}{H\tsub{50}} &
    \multicolumn{1}{c}{H\tsub{1}} \\
    \hline
NEA      &  16062 & $1.9_{-0.6}^{+1.1} \times 10^{5}$ & $7.2 \pm 0.4$ & $0.8 \pm 0.1$ & $1.4_{-0.5}^{+1.0} \times 10^{4}$ &$1.5_{-0.6}^{+1.0} \times 10^{4}$  & 22.75 & 23.75 & 26.50 \\
MC       &  15488 & $1.2_{-0.8}^{+1.6} \times 10^{5}$ & $9.0 \pm 0.6$ & $0.6 \pm 0.1$ & $1.0_{-0.8}^{+1.7} \times 10^{4}$ &$1.2_{-0.8}^{+1.7} \times 10^{4}$  & 21.00 & 21.25 & 22.75 \\
MB       & 674981 & $4.3_{-0.9}^{+1.0} \times 10^{6}$ & $1.5 \pm 0.0$ & $0.7 \pm 0.0$ & $8.2_{-2.2}^{+2.5} \times 10^{4}$ &$9.7_{-2.2}^{+2.5} \times 10^{4}$  & 19.50 & 20.00 & 21.25 \\
Trojan   &   6762 & $1.3_{-0.7}^{+0.9} \times 10^{5}$ & $5.1 \pm 1.5$ & $0.5 \pm 0.4$ & $7.1_{-4.9}^{+9.3} \times 10^{3}$ &$7.5_{-5.0}^{+9.5} \times 10^{3}$  & 17.00 & 17.25 & 18.25 \\
Centaur  &    470 & $1.8_{-1.0}^{+1.4} \times 10^{4}$ & $12.2 \pm 0.9$ & $0.6 \pm 0.4$ & $2.2_{-1.4}^{+2.1} \times 10^{3}$ &$2.2_{-1.4}^{+2.1} \times 10^{3}$  & 14.75 & 15.50 & 18.25 \\
KBO      &   2331 & $9.8_{-1.9}^{+2.2} \times 10^{4}$ & $4.9 \pm 0.2$ & $0.6 \pm 0.1$ & $5.3_{-1.3}^{+1.6} \times 10^{3}$ &$5.5_{-1.3}^{+1.6} \times 10^{3}$  &  8.25 &  8.75 & 10.00 \\
Comet    &   1301 & $185.2_{-13.5}^{+15.4}$\textcolor{white}{$\times 10^{0}$} & $19.5 \pm 0.5$ & $1.0 \pm 0.3$ & $21.5_{-3.6}^{+4.2}$\textcolor{white}{$\times 10^{0}$} &$38.2_{-4.3}^{+4.9}$\textcolor{white}{$\times 10^{0}$}  & 18.25 & 19.00 & 22.00 \\
\hline
Total    & 717395 & $4.9_{-1.2}^{+1.4} \times 10^{6}$ & $2.1 \pm 0.1$ & $0.7 \pm 0.0$ & $1.2_{-0.4}^{+0.7} \times 10^{5}$ & $1.4_{-0.4}^{+0.7} \times 10^{5}$ \\
\hline
  \end{tabular}
\end{table*}

  \indent We thus compute the probability density function (PDF) of the
  heliocentric distance of each population.
  For that, we compute the 2-D distribution of the semi-major axis
  \textsl{vs} eccentricity
  of each population using bins of 0.05 in AU and eccentricity.
  For each bin, we compute the PDF of heliocentric distance from
  Kepler's second law. We then sum individual PDF from each bin, normalized by
  the number of SSO in each bin divided by the entire population.\\
  \indent We then combine the distribution of solar elongation from the
  reference survey and the PDF of heliocentric distance of each
  population in Eqs.~\ref{eq:r} and~\ref{eq:phase} to obtain a PDF of the
  (H-V) index (Eq.~\ref{eq:H}). The fraction of populations to be
  observed by Euclid at each magnitude is estimated by multiplying the
  CSD of the synthetic
  populations with the cumulative distribution of the (H-V) index, at
  both ends of Euclid magnitude range (V\tsub{AB}\,=\,\numb{17--24.5}, see the
  dot-dashed lines in Fig.~\ref{fig:sfd}). 
  The number of SSOs observable on the entire celestial sphere
  ($\mathcal{N}$\tsub{S}) can be simply
  read on this graph, and are reported in Table~\ref{tab:num}.
  The difference between synthetic and observed
  population also provides an estimate of the potential number of
  objects to be discovered by Euclid down to V\tsub{AB}\,=\,\numb{24.5}. \\
  \indent We then estimate how many of these objects will be observed
  by Euclid.
  For that, we compute the position of all known SSOs every six months
  for the entire duration of Euclid operations (2020 to 2026) by using
  the Virtual Observatory (VO) web service
  SkyBot 3-D\footnote{\href{http://vo.imcce.fr/webservices/skybot3d/}{http://vo.imcce.fr/webservices/skybot3d/}} 
  \citep{2008-ACM-Berthier}.
  This allows to compute the fraction of known SSOs present within the
  area covered by Euclid surveys ($f_W$, $f_D$, and $f_C$ for the
  \textsl{Wide} and \textsl{Deep} surveys, and calibration frames).
  We report these fraction in Table~\ref{tab:num}, except $f_D$ which is 
  negligible (of the order of 1-10 ppm) due to the low number of known
  SSOs on highly inclined orbits
  \citep[although there is a clear bias against discovering such
    objects in current census of SSOs, see][]{2017-AJ-153-Petit,2017-AA-Mahlke}.
  These figures are roughly independent of the epoch for
  all populations but for the Trojans, which are confined around
  Lagrangian L4 and L5 points on Jupiter's orbit and therefore cover a
  limited range in right ascension at each epoch. \\
  \indent Overall, about \numb{\nObs} SSOs are expected to be observed by
  Euclid, in a size range currently unexplored by large surveys.
  This estimate could be refined once dedicated studies of
  the detection envelop of moving objects will be performed on
  simulated data.
  Euclid could discover thousands of outer solar system objects and
  tens of thousands of sub-kilometric main-belt, Mars-crosser, and
  near-Earth asteroids
  (see typical absolute magnitudes probed by Euclid in
  Table~\ref{tab:num}). 
  Nevertheless, the  
  Large Synoptic Survey Telescope \citep[LSST,][]{2009-Book-LSST} is expected to have
  its scientific first-light in 2021. The LSST will repeatedly
  image the sky down to V$\approx$\numb{24}, over a wide range of
  solar elongations, and will be a major
  discoverer of faint SSOs.
  Assuming a discovery rate of
  10,000 NEAs, 10,000 MCs, 550,000 MBAs,
  30,000 Trojans, 3000 Centaurs, 4000 KBOs, and
  1000 comets per year \citep{2009-Book-LSST}, 
  most of the SSOs potentially available for discovery should 
  be discovered by LSST in the southern hemisphere.
  Exploration of small KBOs in the northern hemisphere will however be
  specific of Euclid.

\section{Specificity of Euclid observations of SSOs\label{sec:detect}}

\begin{figure}[t]
  \centering
  \includegraphics[width=\hsize]{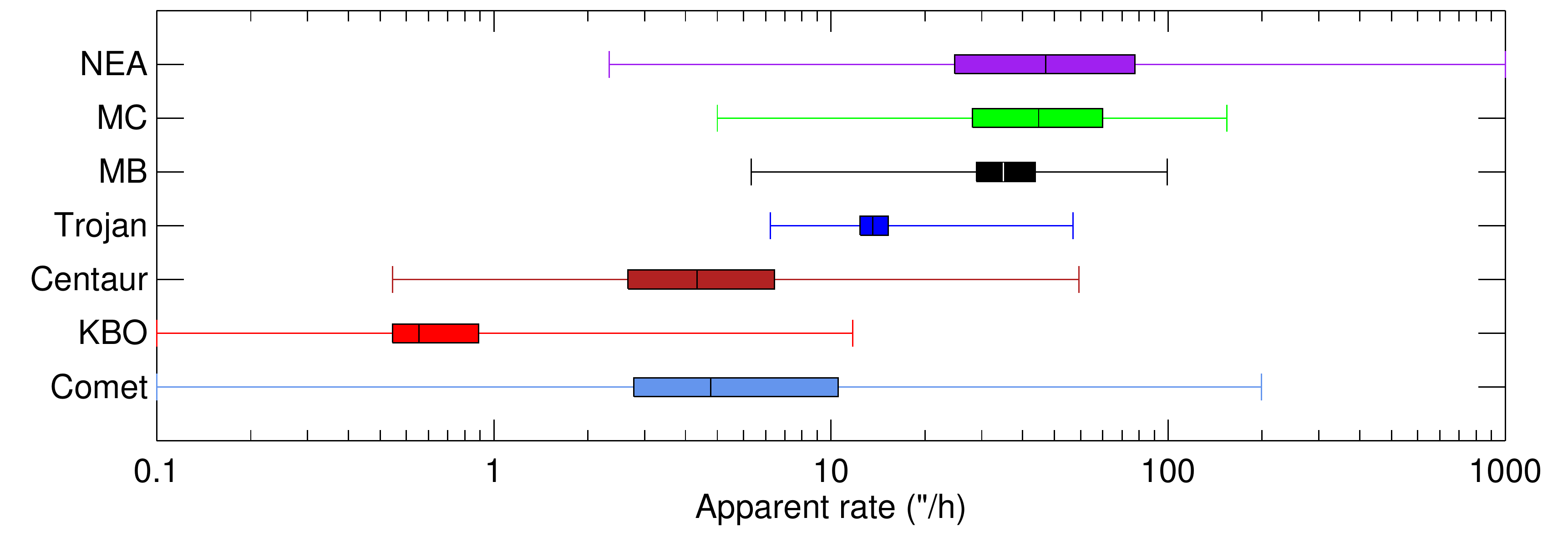}
  \caption[Apparent speed of SSOs]{
    Five-number summary (minimum, maximum, median, 25\% and 75\% quartiles) of
    the apparent rate of each population of 
    SSOs. The Euclid mode of observation at quadrature reduces the
    apparent rate compared to, e.g., opposition.
  }
  \label{fig:rate}
\end{figure}

  \indent The real challenge of SSO observations with Euclid will be
  the astrometry and photometry of highly
  elongated sources (as hinted by Fig.~\ref{fig:img}).
  We present in Fig.~\ref{fig:rate} and Table~\ref{tab:rate} a summary
  of the apparent 
  non-sidereal rate of the different population of SSOs.
  With the exception of the distant-most populations of KBOs,
  Centaurs, and comets, all SSOs will present rates above 10\arcsec/h.
  This implies a motion of hundreds of pixels between the first and last
  VIS frame. 
  During a single exposure, each SSO will move and produce a trailed
  signature, a streak, which length will typically range from 1 to 50 pixels
  for VIS. The situation will be more favorable for NISP, thanks to the
  shorter integration times and larger pixel scale,
  and most SSOs will not trail, or over a few pixels only
  (Table~\ref{tab:rate}).\\
  \indent There have been some recent developments to detect streaks,
  motivated by the optical detection and tacking of artificial
  satellites and debris on low orbits around the Earth.
  Dedicated image processing for trails can be set up 
  to measure the astrometry and photometry of moving objects within 
  a field of fixed stars, without an \textsl{a priori} knowledge of their
  apparent motion \citep[e.g.,][]{2016-AdSpR-57-Virtanen}.
  The success rate in detecting these trails has been shown to reach
  up to \numb{90\%}, even in low signal-to-noise ratio ($\approx$1)
  regime. Such algorithms are currently being tested on simulated
  Euclid data of SSOs (M. Granvik, personal communication).
%
%
%
%
%
\begin{table}[t]
  \caption[Rate and trailing of SSOs observed by Euclid]{
    For each main population of SSOs, we report the
    apparent rate and its 25\% and 75\% quartile variations (i.e.,
    half the population is within these two values),
    and the length of the trails on the detector during the
    simultaneous VIS imaging and NISP spectroscopic \numb{565}\,s
    exposure, and the  
    following NISP Y, J, and H imaging exposures of
    \numb{121}, \numb{116}, and \numb{81}\,s.
  }
  \label{tab:rate}
  \centering
  \begin{tabular}{lrrcccc}
    \hline\hline
Population & Rate & VIS & NISP & Y & J & H \\
& (\arcsec/h) & (pix) & (pix) & (pix) & (pix) & (pix) \\
    \hline
NEA        & 43.3$_{\scriptscriptstyle -19.9}^{\scriptscriptstyle +36.5}$ &  67.9 &  22.6 &   4.8 &   4.6 &   3.2 \\
MC         & 41.3$_{\scriptscriptstyle -14.9}^{\scriptscriptstyle +22.6}$ &  64.8 &  21.6 &   4.6 &   4.4 &   3.1 \\
MB         & 32.5$_{\scriptscriptstyle  -5.5}^{\scriptscriptstyle  +7.9}$ &  51.0 &  17.0 &   3.6 &   3.5 &   2.4 \\
Trojan     & 13.3$_{\scriptscriptstyle  -1.1}^{\scriptscriptstyle  +1.4}$ &  20.9 &   7.0 &   1.5 &   1.4 &   1.0 \\
Centaur    &  4.0$_{\scriptscriptstyle  -1.5}^{\scriptscriptstyle  +2.9}$ &   6.2 &   2.1 &   0.4 &   0.4 &   0.3 \\
KBO        &  0.6$_{\scriptscriptstyle  -0.1}^{\scriptscriptstyle  +0.3}$ &   1.0 &   0.3 &   0.1 &   0.1 &   0.0 \\
Comet      &  4.4$_{\scriptscriptstyle  -1.8}^{\scriptscriptstyle  +6.2}$ &   6.9 &   2.3 &   0.5 &   0.5 &   0.3 \\
    \hline
  \end{tabular}
\end{table}
%
%

\section{Source identification, astrometry, and dynamics\label{sec:astro}}
  \indent As established in Section~\ref{sec:disco}, Euclid will
  observe of the order of \numb{\nObs} SSOs, even if its nominal
  survey is avoiding ecliptic latitudes below \numb{15}\degr, with the
  notable exception of the calibration fields (Fig.~\ref{fig:rs}).\\
  \indent The design of the surveys, with hour-long sequences of
  observation of each field, will however preclude orbit determination
  for newly discovered objects. 
  This hour-long coverage is nevertheless sufficient to discriminate
  between NEAs, MBAs, and KBOs \citep{2017-AA-Spoto}.
  The situation will be very similar to the SDSS Moving Object
  Catalog (MOC), in which many SSO sightings corresponded to
  unknown objects at the time of the release \citep[still about 53\%
    at the time of the 4\tup{th}
    release,][]{2001-AJ-122-Ivezic,2002-AJ-124-Ivezic}. 
  Attempts for identification will have to be regularly performed \textsl{a
  posteriori} once the number of known objects, hence orbits, will
  increase, like we did for the SDSS MOC, identifying
  27\% of the unknown sources \citep[][]{2016-Icarus-268-Carry},
  using the SkyBoT Virtual Observatory tool
  \citep[][]{2006-ASPC-351-Berthier, 2016-MNRAS-458-Berthier}.
  The success rate for \textsl{a posteriori} identification of SSOs
  detected by Euclid should even be higher than in aforementioned
  study, as the LSST will be 
  sensitive to the same apparent magnitude range. \\
  \indent Compared with tens of points over
  many years provided by the LSST, the astrometry by Euclid should
  contribute little to the determination of SSO orbits, with the
  following exceptions.
  First, the objects in the outer solar system (Centaurs and KBOs) in
  the northern hemisphere will not be observed by LSST.
  In this respect, the \textsl{Deep} survey will allow to study the
  population of highly inclined Centaurs and KBOs 
  \citep[e.g.,][]{2017-AJ-153-Petit}, thanks to the repeated
  observations of the northern Ecliptic cap (about \numb{40} times). 
  Second, the parallax between the Earth and the Sun-Earth L2 point is
  large, from about a degree 
  for asteroids in the inner belt, to a few tens of arcseconds for
  KBOs. 
  Simultaneous observation of the same field from the two
  locations thus provides the distance of the SSO,
  reducing drastically the possible orbital parameters
  space \citep{2011-CeMDA-109-Eggl}.
  Thus, an interesting synergy between LSST and
  Euclid would reside in planning these 
  simultaneous observations \citep[see,][]{2017--Rhodes}.
  The practical implementation may however be difficult as the
  observations by Euclid at a solar elongation $\Psi$ of
  91.0\,$\pm$\,1.5\degr~impose observations close to
  sunset or sunrise from LSST.

\section{Photometry and spectral classification\label{sec:taxo}}
%
\begin{figure}
  \centering
  \includegraphics[width=\hsize]{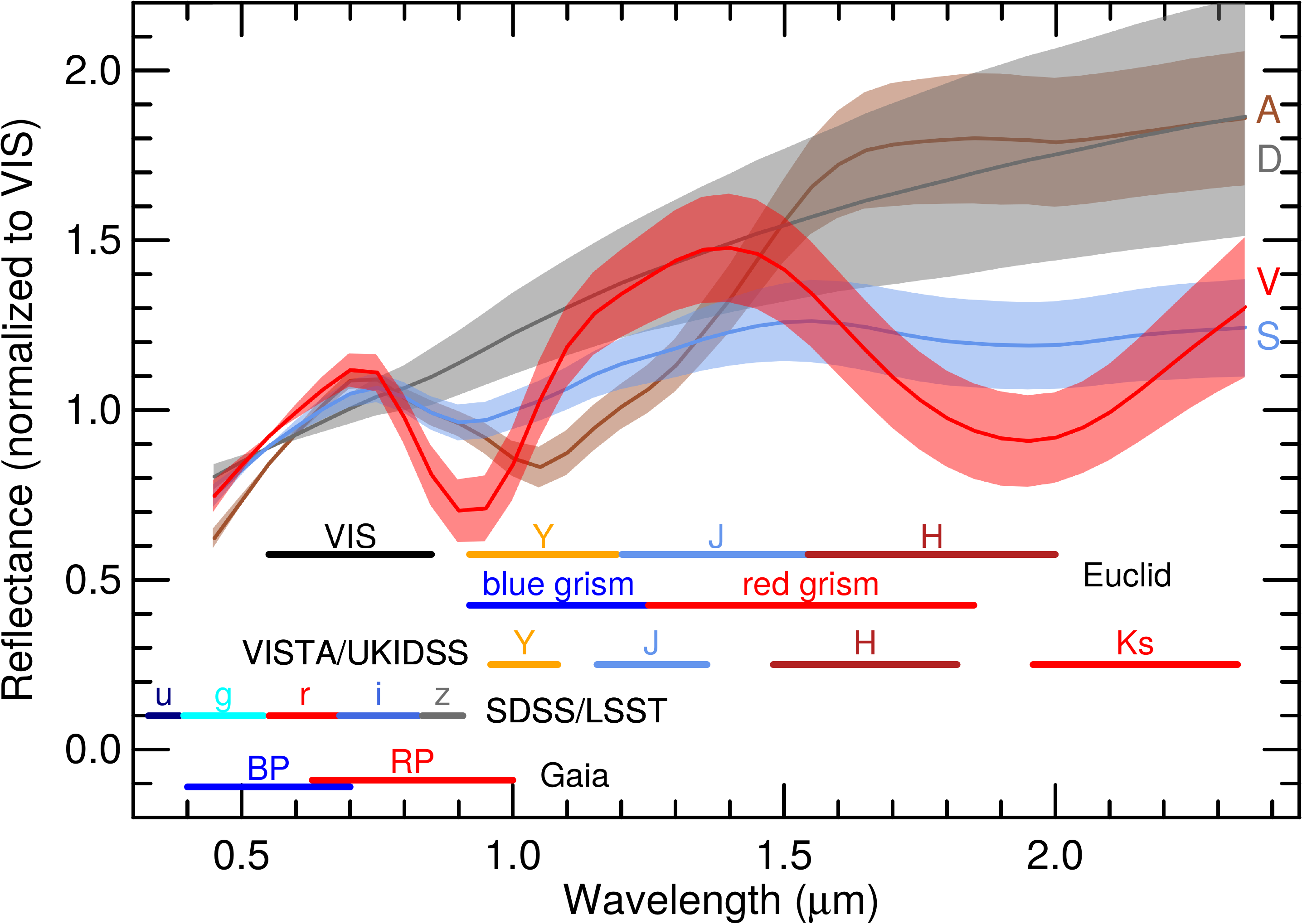}
  \caption[Spectral degeneracy of asteroid classes]{
    Examples of asteroids classes (A, L, S, and V) which are
    degenerated over visible wavelength range.
    For reference, the wavelength
    coverage of each photometric filter and grism on-board Euclid is
    shown, together with 
    SDSS and LSST set of filters
    \citep[u, g, r, i, z,][]{2001-AJ-122-Ivezic},
    2MASS and VISTA, 
    and Gaia blue and
    red photometers (BP, RP) that will produce low-resolution spectra
    \citep[resolving power of a few tens,][]{2012-PSS-73-Delbo}.
  }
  \label{fig:degen}
\end{figure}
%
  \indent In this section we study the impact of Euclid on spectral
  classification of SSOs, thanks to the determination of their spectral
  energy distribution (SED, see Appendix~\ref{app:phot})
  over a large wavelength range, from the
  visible with VIS (\numb{0.5}\,$\mu$m) to the near-infrared with NISP
  (\numb{2}\,$\mu$m). While colors in the visible have been
  and will be obtained for several 10$^{6}$ SSOs thanks to surveys
  like ESA Gaia and the LSST \citep{2016-AA-595-Prusti,
    2009-Book-LSST}, collection of near-infrared photometry is
  lacking. 
  The only facility currently operating from which  
  near-infrared colors for numerous SSOs have been obtained is
  the ESO VISTA telescope \citep{2016-AA-591-Popescu}.
  As described above, the upcoming 
  ESA Euclid \citep[and also the NASA WFIRST mission which
    shares many specifications with Euclid, see][]{2012-arxiv-Green, 2017-arXiv-Holler} may
  radically change 
  this situation.\\
  \indent At first order, SSOs display a G2V spectrum at optical wavelength,
  due to the reflection of the Sun light by their surface.
  Depending on their surface composition, regolith packing, and degree
  of space weathering, their
  spectra are however modulated by absorption bands and slope effects.
  Historically, SSOs spectra have always been studied in
  \textsl{reflectance}, that is their recorded spectrum divided by the 
  spectrum of the Sun, approximated by a G2V star observed with the same
  instrument setting as the scientific target.
  The colors and low-resolution (R\,$\approx$\,\numb{300-500}) of
  asteroids have been used since decades to classify them, in a scheme
  called taxonomy, using the visible range only, or the near-infrared
  only, or both 
  \citep[see,][]{1975-Icarus-25-Chapman,1987-Icarus-72-Barucci,
    2002-Icarus-158-BusI,
    2002-Icarus-158-BusII,2009-Icarus-202-DeMeo}.
  For KBOs, broad-band colors and medium-resolution
  (R$\approx$3000--5000) have been used to characterize their
  surface composition
  \citep[e.g.,][]{2010-AA-511-Snodgrass,2011-AA-534-Carry, 2012-AA-544-Carry},
  although current taxonomy is based on 
  broad-band colors only \citep{2008-SSBN-3-Fulchignoni}.
  \\
\begin{figure}[t]
  \centering
  \includegraphics[width=\hsize]{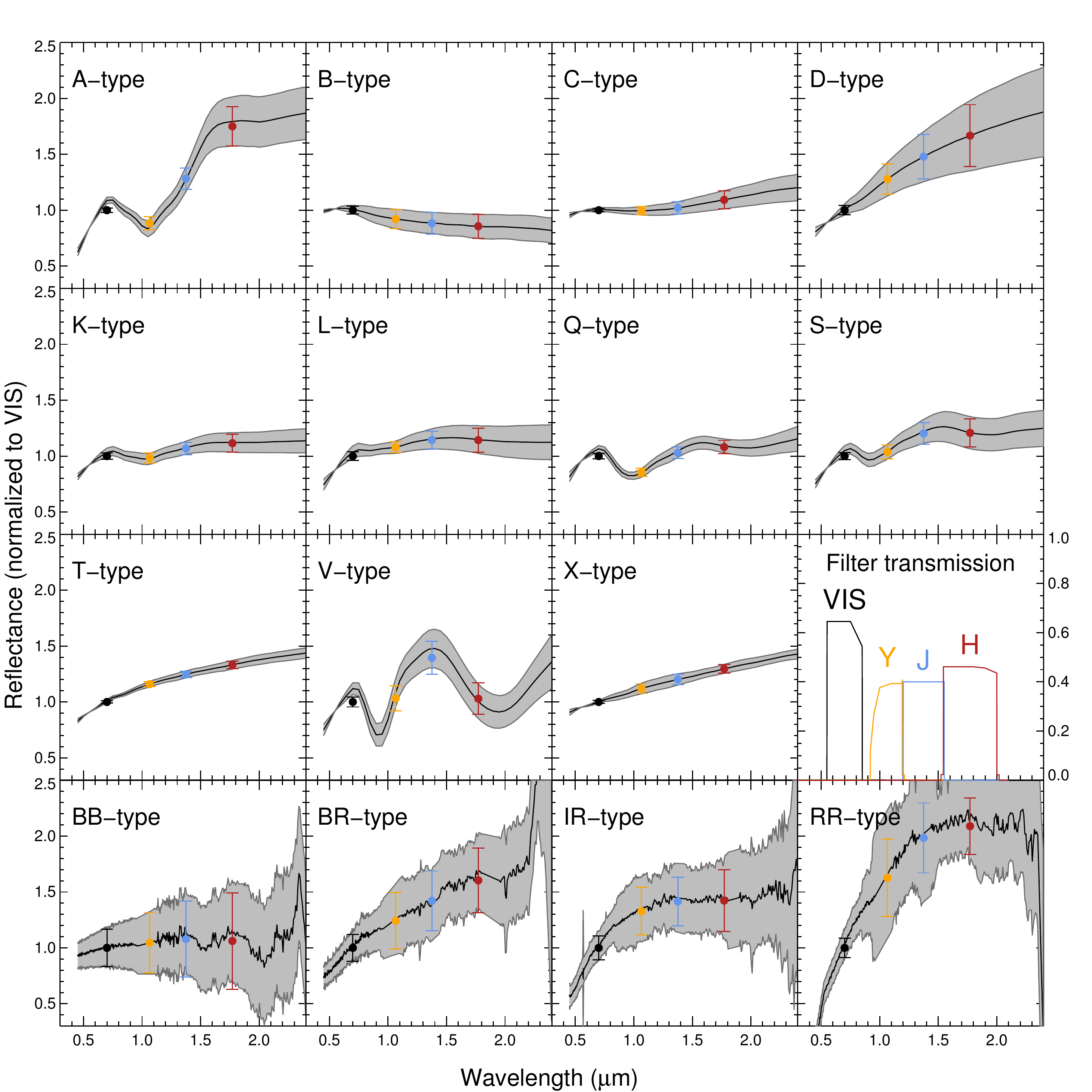}
  \caption[Asteroid taxonomy into Euclid photometry]{
    The eleven asteroid (A- to X-type) and
    four KBO (BB, BR, IR, RR) spectral classes considered here, converted
    into photometry for classification simulation (see text).
    The transmission curves of VIS and NISP filters are also plotted
    for reference.
  }
  \label{fig:demeo2phot}
\end{figure}
%
  \indent Information on the taxonomic class has been derived for
  about 4000 asteroids based on their low-resolution spectra
  \citep[mainly from SMASS, SMASSII, and S$^3$OS$^2$ surveys, see][]{
  2002-Icarus-158-BusI, 2002-Icarus-158-BusII,
  2004-Icarus-172-Lazzaro}.
  Using the broad-band photometry from the Sloan Digital Sky Survey (SDSS),
  many studies have classified tens of thousands of asteroids
  \citep[e.g.,][]{2001-AJ-122-Ivezic, 2002-AJ-124-Ivezic,
    2005-Icarus-173-Nesvorny, 2010-AA-510-Carvano,
    2013-Icarus-226-DeMeo}. These studies opened a new era in the
  study of asteroid families \citep{2013-MNRAS-433-Carruba}, 
  space weathering \citep{2005-Icarus-173-Nesvorny, 2012-Icarus-219-Thomas},
  distribution of material in the inner solar system
  \citep{2014-Nature-505-DeMeo, 2014-Icarus-229-DeMeo}, 
  and origins of near-Earth asteroids \citep{2016-Icarus-268-Carry}.
  The on-going survey by ESA Gaia will provide low-resolution
  spectra (R$\approx$35) for 300,000 asteroids, with high photometric
  accuracy, and the taxonomic class will be determined for
  each SSO \citep{2012-PSS-73-Delbo}. \\
  \indent Nevertheless, any classification based on SDSS,
  Gaia, or LSST (which will use a filter set comparable with SDSS),
  suffers from a wavelength range limited to the visible only.
  It is, however, known that several classes are degenerated over this
  spectral range, and only near-infrared colors/spectra can
  disentangle them \citep[Fig.~\ref{fig:degen} and][]{2009-Icarus-202-DeMeo}.
  The near-infrared photometry provided by Euclid will therefore be
  highly valuable, alike that reported from 
  2MASS \citep{2000-Icarus-146-Sykes}
  or ESO VISTA VHS \citep{2013-Messenger-154-McMahon,
    2016-AA-591-Popescu} surveys.\\  
  \indent To estimate the potential of Euclid photometry for spectral
  classification of asteroids, we simulate data using the visible and near-infrared
  spectra of the 371 
  asteroids that were used to create the Bus-DeMeo taxonomy
  \citep{2009-Icarus-202-DeMeo}, and of \numb{43} KBOs with known
  taxonomy \citep{2017-AA-604-Merlin}.
  We convert their reflectance spectra into photometry
  (Fig.~\ref{fig:demeo2phot}), taking the 
  reference VIS and NISP filter transmission curves\,\footnote{Available on
    Geneva university 
  \href{http://www.isdc.unige.ch/euclid/total-transmission-filter}{Euclid
    pages}}.\\
\begin{figure}[t]
  \centering
  \includegraphics[width=\hsize]{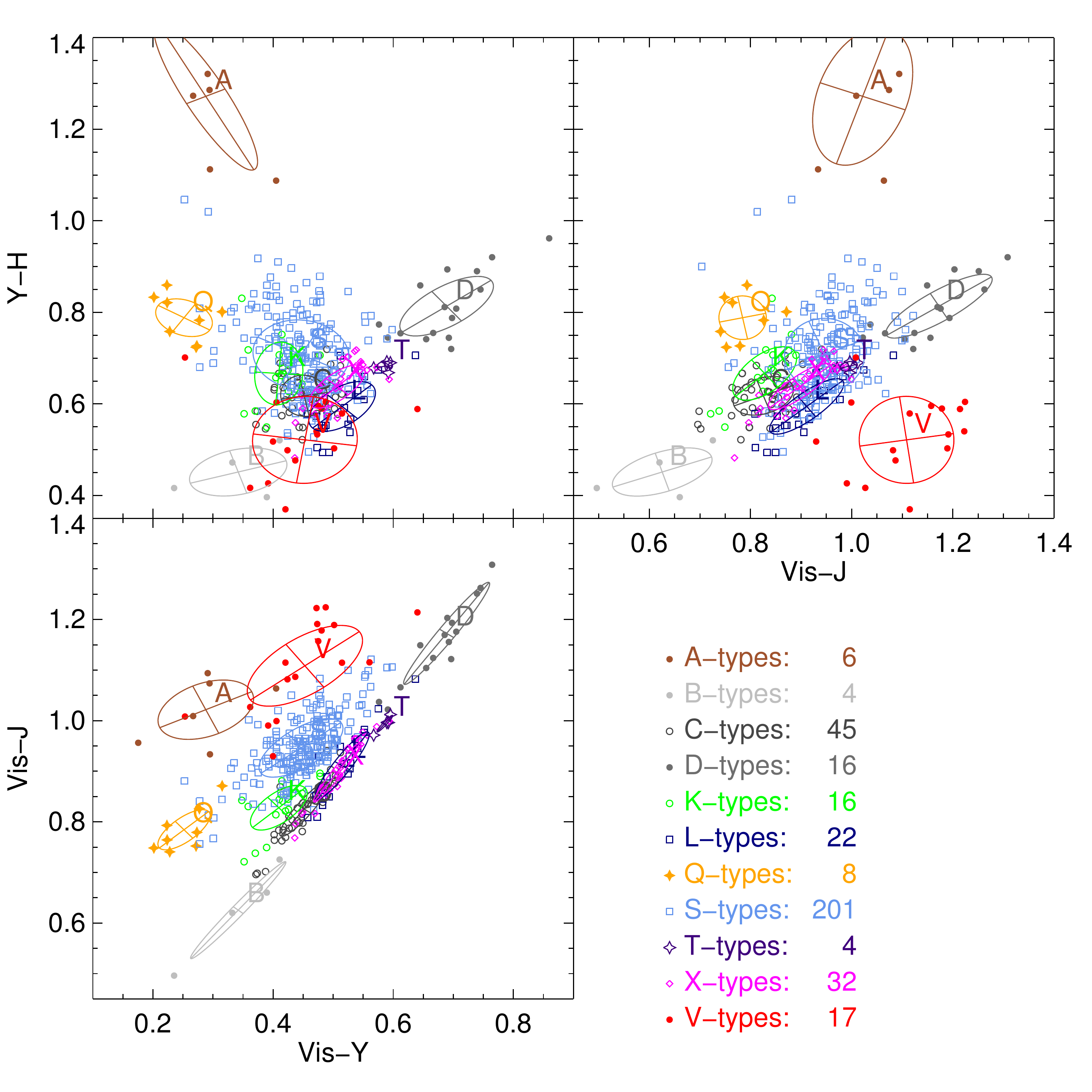}
  \caption[Asteroid classification]{
    Result of the classification of the 371 asteroids from Bus-DeMeo
    taxonomy, presented in three filter combinations: 
    VIS-Y, VIS-J, and Y-H.
    Several extreme classes, like A, B, D, V, and T,
    can be easily sorted out thanks to the large wavelength coverage
    of Euclid.
  }
  \label{fig:PCA}
\end{figure}
%
  \indent One key aspect of Euclid operations for determining the
  colors of SSO is the repetition of the four-filters sequence over an
  hour. Thus,   each filter will be bracketed by other filters in
  time. This will 
  allow to determine magnitude difference between each pair of filters
  without biases otherwise introduced by the intrinsic 
  variation of the target (Appendix~\ref{app:phot}). For a
  detailed discussion on that effect, see \citet{2016-AA-591-Popescu}.\\
  \indent For each class and combination of filter, we compute the
  average color, dispersion, and co-variance. This allows to classify objects based
  on their distance to all the class centers, normalized by the
  typical spread of the class \citep{2017-PhD-Pajuelo}.
  This learning sample is of course limited in number, and all classes
  are not evenly represented. It nevertheless allows to estimate Euclid
  capabilities by applying the classification scheme to the same
  sample. This is presented in Fig.~\ref{fig:PCA}.
  The leverage provided by the long wavelength coverage allows to
  clearly identify several classes: A, B, D, V, Q, and T
  \citep{2009-Icarus-202-DeMeo}. The main classes in the asteroid
  belt, the C, S, and X \citep{2014-Nature-505-DeMeo}, are more
  clumped, and our capabilities to classify them will 
  depend on the exact throughput of Euclid optical path.
%
%
\begin{figure}[t]
  \centering
  \includegraphics[width=\hsize]{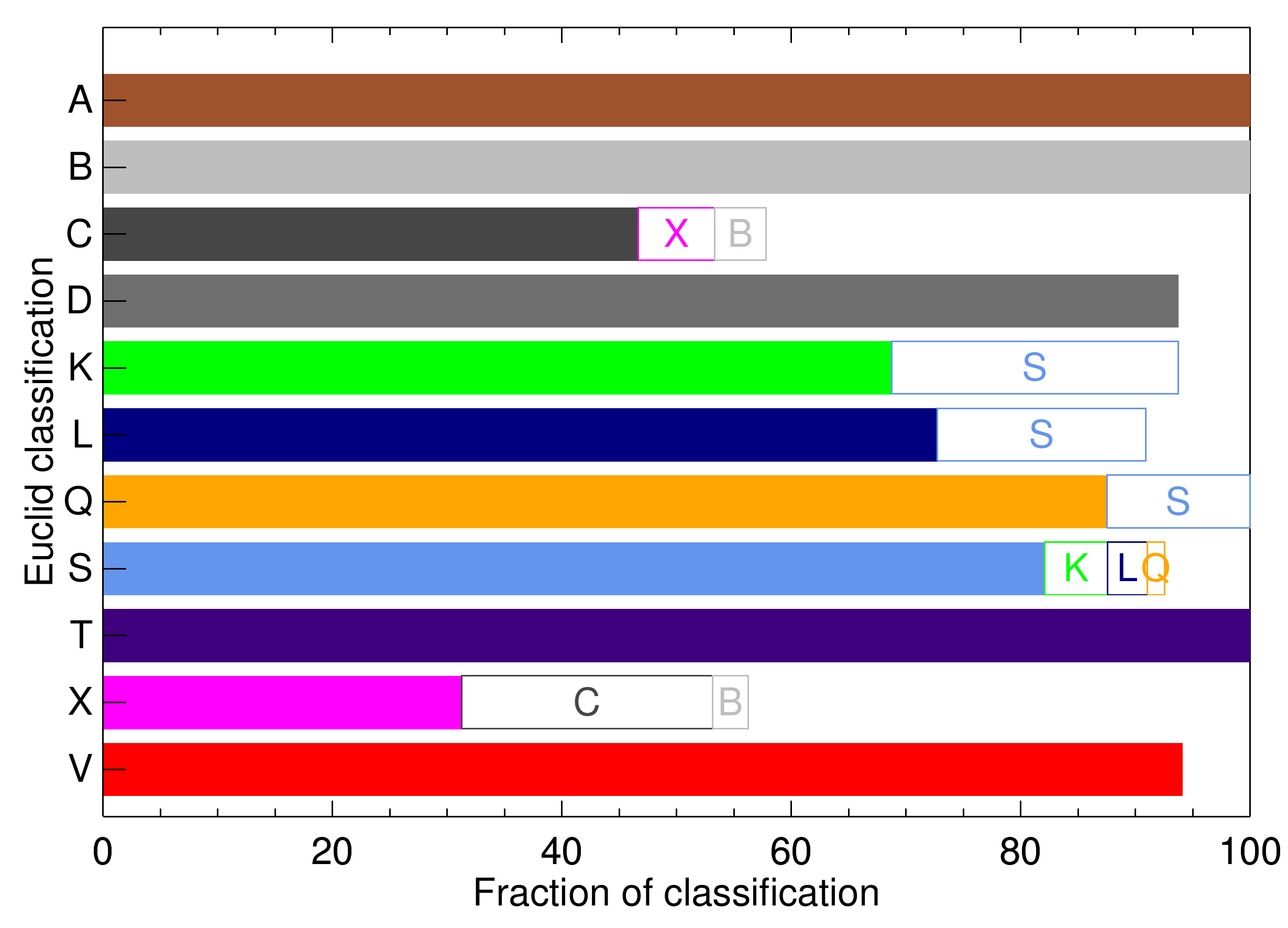}
  \caption[Validation of asteroid classification]{
    Percentage of correct (solid bar) and compatible (open bar)
    classification for each Bus-DeMeo taxonomic classes.
    The Euclid photometry alone allows to classify asteroids into 11
    classes.
  }
  \label{fig:taxo}
\end{figure}
%
  \indent For KBOs, their spectral behavior from the blue-ish BB to the
  extremely red RR will place them in these graphs along a line
  going though the C, T, and D-types (which colors are close to the BB,
  BR, and IR classes). The RR-types will be even further from the
  central clump than the D-types.
  Identifying the different KBO spectral classes should therefore be
  straightforward with Euclid set of filters. \\
  \indent In all cases, spectral characterization using Euclid colors
  will benefit from the colors and spectra in the visible observed by
  Gaia and LSST \citep{2012-PSS-73-Delbo, 2009-Book-LSST}, 
  visible albedo \citep[from IRAS, AKARI, WISE, Herschel
    observations, e.g.,][]{2002-AJ-123-Tedesco-a, 2009-EMP-105-Muller,
    2011-ApJ-741-Masiero, 2011-PASJ-63-Usui}, and
  solar phase function parameters \citep[see][for
    an example of the use of phase function for
    taxonomy]{2012-Icarus-219-Oszkiewicz}. 
  The success rate of classification from Euclid photometry only hence
  represents a lower estimate.\\ 
  \indent We present in Fig.~\ref{fig:taxo} the success rate of
  classification of the 371 asteroids from the Bus-DeMeo taxonomy.
  The classes are generally recovered with a success rate above 60\%,
  and when misclassified, asteroids end up in spectrally similar
  (compatible) classes with a success rate closer to 90\%, but
  for the C and X classes. 
  We do not repeat the exercise for KBOs given the limited size of the
  available sample. Their spectral classes being, however, much alike the C, T, and
  D-type asteroid, and even redder, their identification should be
  straightforward with Euclid filter set.\\
  \indent In summary, the VIS and NISP photometry that will be
  measured by Euclid 
  seems very promising to class SSOs among their historical spectral
  classes.

\section{Near-infrared spectroscopy with NISP\label{sec:spec}}
  \indent Euclid will also acquire near-infrared low-resolution
  (resolving power of \numb{380}) spectra for many SSOs, down to
  m$_{AB}$\,$\approx$\,21, i.e., similar to Gaia limiting magnitude. 
  Simultaneously to the four VIS exposures, NISP will acquire four
  slitless spectra of the same field of view.
  In the \textsl{wide} survey, only the \textsl{red} grism
  (1.25 to 1.85\,$\mu$m) will be used, the usage of the \textsl{blue} grism (0.92
  to 1.25 $\mu$m) being limited to the \textsl{deep} survey.
  The \textsl{red} grism will cover typical absorption bands
  of volatile compounds (e.g., water or methane ices) such as found on
  distant KBOs. The main diagnostic features of asteroids (NEAs, MBAs) are
  however located within the blue arm at 1\,$\mu$m, and
  at 2\,$\mu$m, outside the spectral range of the \textsl{red} grism. \\
  \indent Because there is no slit, many sources will be blended. 
  To decontaminate each slitless spectrum from surrounding sources,
  the exposures will be taken with three different grism orientations, 
  90\degr~apart.
  For exposures with the spectral dispersion aligned with the ecliptic, i.e., parallel to the typical
  SSO motion, as each SSO will blend with itself.
  For the remaining orientations, SSOs will often blend with background
  sources, degrading both spectra. This may be an issue for the \textsl{wide}
  survey in its lowermost ecliptic latitude range, where many sources
  will be blended with G2V spectra from SSOs.\\
  \indent The apparent motion of outer solar
  system objects being limited (Table~\ref{tab:rate}), their spectra
  may be extracted by the Euclid consortium tools, designed to work on
  elongated sources (typically 1\arcsec). Near-infrared spectra for
  thousands of Centaurs and KBOs could thus be produced by Euclid.  
  For objects in the
  inner solar system, the extraction of
  their spectra may be challenging, and in-depth assessment of the
  feasibility of such measurements is beyond the scope of this paper.
  In both cases, these spectra be very similar to the
  low-resolution spectra used to define current asteroid taxonomy
  \citep{2009-Icarus-202-DeMeo} and diagnostic of KBOs class as
  defined by \citep{2008-SSBN-3-Fulchignoni}.

\section{Multiplicity and activity of SSOs\label{sec:psf}}
  \indent With a very stable PSF and a pixel scale of 0.1\arcsec~and
  0.3\arcsec~for VIS and NISP,
  close to the diffraction limit of Euclid,
  the morphology of sources can be studied. This is indeed one of the
  main goals of the cosmological survey \citep{2011-Euclid-Laureijs}.
  We first assess how Euclid could detect satellites around SSOs, and
  then activity, i.e., dust trails.

  \subsection{Direct imaging of multiple systems with Euclid\label{sec:bin}}
    \indent In two decades since the discovery of the first satellite
    of asteroids, Dactyl around (243) Ida, by the Galileo mission
    \citep{1995-Nature-374-Chapman}, direct imaging has been the main
    source of discovery and characterization of satellites around
    large SSOs, in the main belt \citep[e.g.,][]{1999-Nature-401-Merline,
      2014-Icarus-239-Berthier}, among 
    Jupiter Trojans \citep{2006-Nature-439-Marchis,
      2014-ApJ-783-Marchis}, 
    and KBOs \citep[e.g.,][]{2005-ApJ-632-Brown, 2006-ApJ-639-Brown,
      2010-AJ-139-Brown, 2011-AA-534-Carry, 2017-NatAs-1-Fraser}.  
    This is particularly evident for KBOs, for which
    65 of the 80 known binary systems where discovered by the
    Hubble Space Telescope, and the other 14 by large ground-based
    telescopes, often supported by adaptive optics
    \citep[see, e.g.,][]{2011-ApJ-743-Parker,
      PDSSBN-Johnston, 2015-AsteroidsIV-Margot}.
    The situation is different for NEAs and small MBAs, for
    which most discoveries and follow-up observations were made with
    optical lightcurves and radar echoes
    \citep[e.g.,][]{2007-Icarus-190-Pravec, 2012-Icarus-218-Pravec,
      2011-AJ-141-Fang, 2011-Icarus-216-Brozovic}. \\
    \indent To estimate Euclid capabilities to angularly resolve a
    multiple system, we use the compilation of system parameters by
    \citet{PDSSBN-Johnston}. We compute the magnitude difference
    between components $\Delta m$ from their diameter ratio, and their
    typical separation $\Theta$ from the ratio of the binary system
    semi-major axis to its heliocentric semi-major axis
    (Table~\ref{tab:bin}). \\
%
\begin{table}[!t]
  \caption[Properties of multiple systems per population]{
    Typical magnitude difference ($\Delta m$) and
    angular separation ($\Theta$) between components of multiple SSO
    systems.
    NEA and MCs share similar characteristics, and so does large MBAs
    and Trojans. We split MBAs into two categories, according to the
    diameter $D$
    of the main component.
    Estimates on the frequency of binaries in each populations are
    based on the reviews by \citet{2008-SSBN-Noll} and
    \citet{2015-AsteroidsIV-Margot}.
    We only consider high-inclination
    KBOs here, the fraction of binaries in the cold belt being closer
    to 30\% \citep{2017-NatAs-1-Fraser}.    
  }
  \label{tab:bin}
\centering
  \begin{tabular}{l rrr}
    \hline
    \hline
    Population & \multicolumn{1}{c}{$\Delta m$} & \multicolumn{1}{c}{$\Theta$}  & \multicolumn{1}{c}{$f$}\\
    & \multicolumn{1}{c}{(mag)} & \multicolumn{1}{c}{(\arcsec)} & \multicolumn{1}{c}{(\%)} \\
    \hline
    NEA \& MC          & $1.8_{-1.8}^{+2.0}$ & $0.01_{-0.01}^{+0.01}$ & $15 \pm 5$ \\
    MBA ($D<10$\,km)   & $2.5_{-0.9}^{+0.9}$ & $0.01_{-0.01}^{+0.01}$ & $15 \pm 5$ \\
    MBA ($D>100$\,km)  & $5.4_{-2.7}^{+2.7}$ & $0.30_{-0.25}^{+0.25}$ & $3 \pm 2$  \\
    KBO                & $1.5_{-1.5}^{+2.0}$ & $0.43_{-0.43}^{+0.60}$ & $6 \pm 4$  \\
    \hline
  \end{tabular}
\end{table}
%
%
  \indent The angular resolution of Euclid will thus allow to detect
  satellites of KBOs and large MBAs, but not those around NEAs, MCs,
  and small MBAs.
  The case of KBOs is straightforward, owing to the very little
  smearing of their PSF from their apparent motion
  (Table~\ref{tab:rate}). 
  Based on the expected number of observations of KBOs
  (Table~\ref{tab:num}) and their binarity fraction, Euclid should
  observe 300\,$\pm$\,200 multiple KBO systems, i.e., a four fold
  increase. \\ 
  \indent The case of MBAs is more complex.
  First, there are only 25 large MBAs with an inclination higher than
  15\degr, i.e., potentially observable by Euclid.
  Second, the fraction and properties of multiple systems
  for MBAs with a diameter between 10 and 100\,km is 
  \textsl{terra incognita}. This is due to observational biases: 
  detection by lightcurves is more efficient on close-by components,
  and direct imaging, especially from ground-based telescopes using
  adaptive optics, focused on bright, hence large, primaries.
  If most binaries around small asteroids ($D<10$\,km) are likely
  formed by rotational fission caused by YORP spin-up
  \citep{2008-Nature-454-Walsh, 2010-Nature-466-Pravec,
    2015-AsteroidsIV-Walsh},
  satellites of larger bodies are the result of
  re-accumulation of ejecta material after impacts
  \citep{2001-Science-294-Michel, 2004-Icarus-170-Durda}. Some
  satellites around mid-sized MBAs are therefore to be expected, but
  with unknown frequency.
  Considering a ratio of $\approx$5 between the 
  semi-major axis of binary system and the diameter of the main
  component \citep[typical of large MBAs,
    see][]{2015-AsteroidsIV-Margot} and the size distribution of
  high-inclination MBAs,
  only a handful of potential systems would have separations angularly
  resolvable by Euclid. 
  Finally, the apparent motion of MBAs implies highly elongated PSFs,
  diminishing even further the fraction of detectable systems.\\
  \indent For these reasons, Euclid will therefore contribute
  little, if at all, to the characterization of multiple systems among
  asteroids. Prospects for discoveries of KBO binaries is however very
  promising.

  \subsection{Detection of activity\label{sec:activity}}
    \indent The distinction between comets and other kind of small
    bodies in our Solar System is, by convention, based on the
    detection of activity, i.e., of unbound atmosphere
    also called coma. Comets cannot be distinguished from their
    orbital elements only (Fig.~\ref{fig:pop}).
    The figure blurred further with the discovery of comae around
    Centaurs, and even MBAs, called active asteroids \citep[see][for
      reviews]{2009-AJ-137-Jewitt, 2015-AsteroidsIV-Jewitt}. \\
    \indent The cometary-like behavior of these objects was discovered 
    either by sudden surges in magnitude, or diffuse non-point-like
    emission around them. There are currently 18 known active
    asteroids and 12 known active Centaurs, corresponding to 
    25 ppm and 13\% of their host populations respectively.
    The property of the observed comae is typically 1 to 5 magnitude
    fainter than the nucleus, within a 3\arcsec~radius 
    \citep[although this large aperture was chosen to avoid
      contamination from the nucleus PSF which extended to about
      2\arcsec~due to atmospheric seeing,][]{2009-AJ-137-Jewitt}.\\
    \indent With much higher angular resolution, and its very stable
    PSF required for its primary science goal
    \citep{2011-Euclid-Laureijs}, Euclid has the capability to detect such
    activity. Based on the expected number of observations
    (Table~\ref{tab:num}) and the aforementioned fraction of observed
    activity, 
    Euclid could observe a couple of active asteroids and about
    300$_{-200}^{+300}$ active Centaurs.
    As in the case of multiple systems however,
    detection capability will be
    diminished by the trailed appearance of SSOs.
    This will be dramatic for MBAs, but limited for Centaurs
    (Table~\ref{tab:rate}): typical motion will be of 6 pixels, i.e.,
    0.6\arcsec, while typical coma extend over several arcseconds.

\section{Time-resolved photometry\label{sec:photo}}
\begin{figure}[t]
  \centering
  \includegraphics[width=.9\hsize]{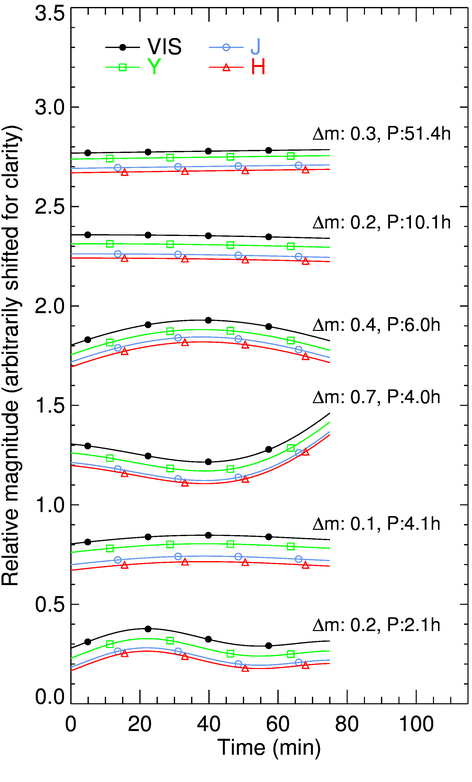}
  \caption[Simulated multi-filter lightcurves with Euclid]{
    Examples of simulated SSO multi-filter lightcurves as observed by
    Euclid VIS and NISP. For each lightcurve, the amplitude ($\Delta m$) and
    rotation period (P) is reported.
    For each, the four lightcurves corresponding to the different filters
    are printed (with a magnitude difference reduced by a factor 10
    for clarity), together with the photometry at the cadence of Euclid.
  }
  \label{fig:lc}
\end{figure}

  \indent The observations of each field, in four repeated sequences of
  VIS and NISP photometry, will provide hour-long lightcurves sampled
  by 4\,$\times$\,4 measurements, or a single lightcurve made of 16
  measurements by converting all magnitudes from
  the knowledge of the SED (Fig.~\ref{fig:lc}, Appendix~\ref{app:phot}). \\
%
%
%
  \indent Since decades, optical lightcurves have been the prime data
  set for 3-D shape modeling 
  and study of SSO multiplicity from mutual eclipses
  \citep[see the reviews by][]{2015-AsteroidsIV-Margot,
    2015-AsteroidsIV-Durech}. 
  Taken alone, a single lightcurve, such as those Euclid will provide,
  does not provide much constrains. Both shape and dynamical modeling
  indeed require multiple Sun-target-observer geometries, which can
  only be achieved by accumulating data over many years and
  oppositions. \\
  \subsection{Period, spin, and 3-D shape modeling}
    \indent Traditionally, the period, spin orientation,
    and 3-D shape of asteroids were determined by using many
    lightcurves taken over several apparitions
    \citep[e.g.,][]{2001-Icarus-153-Kaasalainen-a,2001-Icarus-153-Kaasalainen-b}.
    It has been show later on that photometry measurements, \textsl{sparse} in time\footnote{We call \textsl{sparse photometry}
      lightcurves for which the sampling is typically larger than the
      period, as opposed to \textsl{dense} lightcurves, in which the
      period is sampled by many measurements \citep[see, e.g.,][]{2016-AA-586-Hanus}.},
    convey the
    same information and can be use alone or in combination with \textsl{dense}
    lightcurves \citep{2004-AA-422-Kaasalainen}. Large surveys
    such as Gaia and the LSST will deliver sparse photometry for
    several 10$^{5-6}$ SSOs \citep{2007-EMP-101-Mignard, 2009-Book-LSST}.\\
    \indent In assessing the impact of PanSTARRS and Gaia data on shape
    modeling, \citet{2005-EMP-97-Durech} and \citet{2012-PSS-73-Hanus}
    however showed that searching for the rotation period with sparse
    photometry only may result in many ambiguous solutions.
    The addition of a single dense lightcurve often removes many aliases
    and harmonics in a periodogram, removing the ambiguous solutions, 
    the impact of the single lightcurve depending on the fraction of the
    period it covers (J. Durech, personnal communication). \\
%
\begin{figure}
  \centering
  \includegraphics[width=.9\hsize]{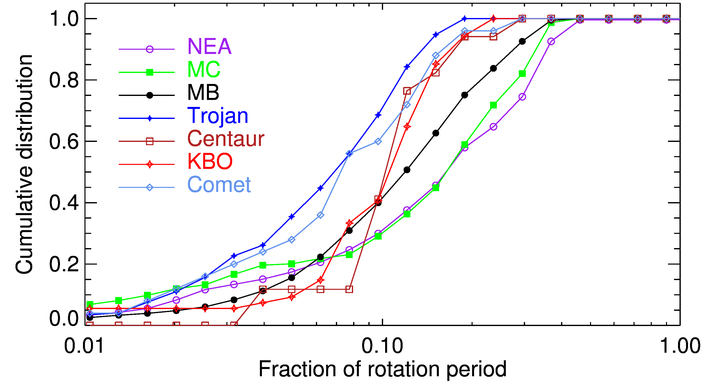}
  \caption[XXX]{
    Cumulative distribution of the fraction of rotation covered by one
    hour of observations, computed on the 5759
    entries with a quality code 2 or 3 from the Planetary Data System archive
    \citep[][]{2017-PDS-Harris}, and the 25 comets
    from \citet{2004-CometsII-Samarasinha} and \citet{2012-AA-548-Lowry}.
  }
  \label{fig:period}
\end{figure}
%
    \indent The rotation periods of SSOs range from a few minutes 
    to several 
    hundreds of hours. The bulk of the distribution is however confined
    between \numb{2.5}\,h  
    \citep[which is called the \textsl{spin barrier}, see
      e.g.,][]{2015-AsteroidsIV-Scheeres} and 10--15\,h.
    This implies that Euclid lightcurves will typically cover
    between \numb{5--10 and 40}\% of the rotation period of SSOs
    (Fig.~\ref{fig:period}). 
    Euclid lightcurves will cover more
    than a quarter of rotation (the maximum
    change in geometry over a rotation, used here as a baseline)
    for
    35\% of NEAs, 28\% of MCs, and 16\% of MBAs, and only a handful of
    outer solar system objects
    The hour-long lightcurves provided by Euclid will thus be valuable
    for 3-D shape modeling of thousands of asteroids 
    ($5.25_{-2.10}^{+3.50} \times 10^{3}$ NEAs, 
    $3.36_{-2.24}^{+4.76} \times 10^{3}$ MCS, and
    $1.55_{-0.35}^{+0.40} \times 10^{4}$ MBAs). 

  \subsection{Mutual events and multiplicity}
    \indent Binary asteroids represent about
    \numb{15\,$\pm$\,5}\% of the population of NEAs larger than 300\,m
    \citep[Sect.~\ref{sec:psf},][]{2006-Icarus-181-Pravec}, and a similar
    fraction is expected among MCs and MBAs with a diameter smaller
    than 10\,km
    \citep[Table~\ref{tab:bin},][]{2015-AsteroidsIV-Margot}.  
    Most of these multiple systems were discovered by lightcurve observations 
    recording mutual eclipsing and occulting events
    \citep[140 of the 205 binary asteroid systems known to date, the
      remaining being mostly binary NEAs discovered by radar echoes,
      see][]{PDSSBN-Johnston}. \\
    \indent These systems have orbital periods of
    \numb{24 $\pm$ 10}\,h, and diameter ratio of
    $0.33 \pm 0.17$, which implies a 
    magnitude drop of $0.11_{-0.08}^{+0.13}$ during mutual eclipses
    and occultations
    \citep[computed from the compilation of the properties of binary
      systems by][]{PDSSBN-Johnston}.
    The hour-long lightcurves provided by Euclid will thus typically
    cover \numb{4$_{-1}^{+3}$}\% of the orbital period. Considering
    the systems are in mutual events for about 20\% of the orbital
    period at the high phase angle probed by Euclid
    \citep[e.g.,][]{2006-Icarus-181-Pravec, 
      2015-Icarus-248-Carry}, there is a corresponding probability of
    $\approx$(5\,$\pm$\,2)\% to witness mutual events.
    Hence, Euclid could record mutual events for 900$_{-450}^{+700}$
    NEAs, MCs, and MBAs, helping characterizing these systems
    in combination with other photometric data sets such as those
    provided by Gaia and the LSST.

\section{Conclusion\label{sec:conclu}}

  \indent We have explored how the ESA mission Euclid can contribute
  to Solar System science. The operation mode of Euclid is by chance
  well designed for detection and identification of moving objects.
  The deep limiting magnitude (V\tsub{AB}\,$\sim$\,\numb{24.5}) of Euclid and large
  survey coverage (even if avoiding low ecliptic latitude) promise
  about 150,000 observations of solar system objects (SSOs),
  in all dynamical classes, from the near-Earth asteroids to
  the distant Kuiper-belt objects, including comets.\\ 
  \indent The spectral coverage of Euclid photometry, from the visible
  to the near-infrared complements the spectroscopy and photometry
  obtained in the visible only by Gaia and the LSST, allowing 
  spectral classification. The hour-long sequence of observations can
  be used to constrain the rotation period, spin orientation, 3-D
  shape, and multiplicty of SSOs, once combined with the sparse
  photometry of Gaia and LSST.
  The high angular resolution of Euclid should allow the detection of
  several hundreds of satellites around KBOs, and activity for the
  same amount of Centaurs. \\
  \indent The exact number of observations of SSOs, the determination of
  the astrometric, photometric, and spectroscopic precision as function
  of apparent magnitude and rate, and the details of data treatments
  will have to be refined, once the instruments will be fully
  characterized. The exploratory work presented here aims at
  motivating further studies, on each aspect of Euclid observation of SSOs. \\
  \indent 
  In summary, against all odds, a survey explicitly avoiding the
  ecliptic promises great scientific prospects for solar system
  research, which could be delivered as Legacy Science for Euclid.
  A dedicated SSO processing is currently being developed within the
  framework on Euclid data analysis pipeline.
  The main goal of the mission will benefit from this addition, from
  the identification of blended sources
  (e.g., stars, galaxies) with SSOs.\\
  \indent 
  Furthermore, any extension of the survey to lower latitude would
  dramatically 
  increase the figures reported here: there are twice as more SSOs
  for every 3\degr~closer to the ecliptic.
  Any observation at low ecliptic latitude, like
  calibration fields, or during idle time of the main survey or after its
  completion, or dedicated to a Solar System survey
  would
  provide thousands of SSOs each time, allowing to study the
  already-known dark matter of our solar system: the low-albedo minor
  planets.

\begin{acknowledgements}
  Present study made a heavy usage of Virtual Observatory tools
  SkyBoT\,\footnote{SkyBoT: \href{http://vo.imcce.fr/webservices/skybot/}{http://vo.imcce.fr/webservices/skybot/}}
  \citep{2006-ASPC-351-Berthier, 2016-MNRAS-458-Berthier}, 
  SkyBoT 3-D\,\footnote{SkyBoT 3-D: \href{http://vo.imcce.fr/webservices/skybot3d/}{http://vo.imcce.fr/webservices/skybot3d/}}
  \citep{2008-ACM-Berthier}, 
  TOPCAT\,\footnote{TOPCAT:
    \href{http://www.star.bris.ac.uk/~mbt/topcat/}{http://www.star.bris.ac.uk/~mbt/topcat/}}, and
  STILTS\,\footnote{STILTS: \href{http://www.star.bris.ac.uk/~mbt/stilts/}{http://www.star.bris.ac.uk/~mbt/stilts/}}
  \citep{2005-ASPC-Taylor}.
  Thanks to the developers for their development and
  reactivity to my requests, in particular J.~Berthier.
  The present article benefits from many discussions, and comments I
  received, and I would like to thank L.~Maquet and C.~Snodgrass for our
  discussions regarding comet properties, the ESA Euclid group at ESAC
  B.~Altieri,  P.~Gomez, H.~Bouy, and R.~Vavrek for our discussions on
  Euclid and SSO, in particular P.~Gomez for sharing the Reference
  Survey with me. 
  Of course, I wouldn't have had these motivating experiences without
  the support of the ESAC faculty (ESAC-410/2016).
  Thanks to F.~Merlin to have created and shared the KBO average
  spectra for present study.
  Thanks also to R.~Laureijs and T.~M{\"u}ller for their constructive
  comments on an early version of this article, and to S.~Paltani and
  R.~Pello for providing the 
  transmission curves of VIS and NISP filters.
\end{acknowledgements}

\bibliographystyle{aa} 
\bibliography{biblio} 

\appendix
\section{Definition of small body populations\label{app:class}}
  \indent We explicit here the boundaries in orbital elements to
  define the population used thorough the article.
  The boundaries for NEAs classes are taken from 
  \citet{2016-Icarus-268-Carry}, 
  and that of the outer
  solar system from \citet{2008-SSBN-2-Gladman}.

\begin{figure*}
  \centering
  \includegraphics[width=\hsize]{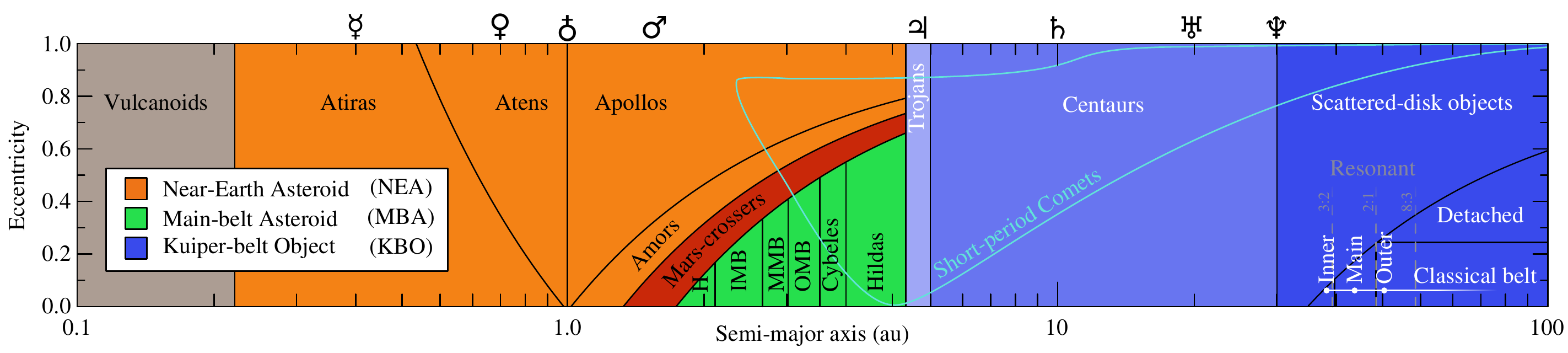}
  \caption[Nomenclature of Solar System Objects]{
    The different classes of SSOs used thorough the
    article. H stands for Hungarias, and IMB, MMB, and OMB for inner,
    middle, and outer belt respectively. 
    \add{Comets orbital elements formally overlap with other classes
      because their classification is based on the presence of a coma
      at short heliocentric distance.}
  }
  \label{fig:pop}
\end{figure*}

\begin{table*}[!t]
\centering
  \begin{tabular}{lrr cc cc cc}
    \hline
    \hline
    Class & \multicolumn{2}{c}{Semi-major axis (au)} & 
            \multicolumn{2}{c}{Eccentricity} & 
            \multicolumn{2}{c}{Perihelion (au)} &
            \multicolumn{2}{c}{Aphelion (au)} \\
          & min. & max. &
            min. & max. &
            min. & max. &
            min. & max.\\
    \hline
%
    NEA                     &  --   &  --     & -- & -- &  --   & 1.300 & -- & -- \\
    \hspace{1em}Atira       &  --   &  a$_\textrm{\earth}$  & -- & -- &  --   &   --  & -- & q$_\textrm{\earth}$ \\
    \hspace{1em}Aten        &  --   &  a$_\textrm{\earth}$  & -- & -- &  --   &   --  & q$_\textrm{\earth}$ & -- \\
    \hspace{1em}Apollo      & a$_\textrm{\earth}$ &  4.600  & -- & -- &  --   & Q$_\textrm{\earth}$ & -- & -- \\
    \hspace{1em}Amor        & a$_\textrm{\earth}$ &  4.600  & -- & -- & Q$_\textrm{\earth}$ & 1.300 & -- & -- \\
    MC                      & 1.300  & 4.600   & -- & -- &1.300  &Q$_\textrm{\mars}$& --& -- \\
    MBA                     & Q$_\textrm{\mars}$ & 4.600 & -- & -- & Q$_\textrm{\mars}$ & -- & -- & -- \\
    \hspace{1em}Hungaria    &  --   & J$_{4:1}$ & -- & -- & Q$_\textrm{\mars}$ & -- & -- & -- \\
    \hspace{1em}IMB         & J$_{4:1}$ & J$_{3:1}$ & -- & -- & Q$_\textrm{\mars}$ & -- & -- & -- \\
    \hspace{1em}MMB         & J$_{3:1}$ & J$_{5:2}$ & -- & -- & Q$_\textrm{\mars}$ & -- & -- & -- \\
    \hspace{1em}OMB         & J$_{5:2}$ & J$_{2:1}$ & -- & -- & Q$_\textrm{\mars}$ & -- & -- & -- \\
    \hspace{1em}Cybele      & J$_{2:1}$ & J$_{5:3}$ & -- & -- & Q$_\textrm{\mars}$ & -- & -- & -- \\
    \hspace{1em}Hilda       & J$_{5:3}$ & 4.600 & -- & -- & Q$_\textrm{\mars}$ & -- & -- & -- \\
    Trojan                  & 4.600 & 5.500 & -- & -- & -- & -- & -- & -- \\
%
    Centaur                 & 5.500 & a$_\textrm{\neptune}$ & -- & -- & -- & -- & -- & -- \\
    KBO                     & a$_\textrm{\neptune}$ &  --    &   --  & -- & -- & -- & -- & -- \\
    \hspace{1em}SDO         & a$_\textrm{\neptune}$ &  --    &   --  & -- & -- & 37.037 & -- & -- \\
    \hspace{1em}Detached    & a$_\textrm{\neptune}$ &  --    & 0.24 & -- & 37.037 & -- & -- & --\\
    \hspace{1em}ICB         & 37.037 & N$_{2:3}$ &   --  & 0.24 & 37.037 & -- & -- & -- \\
    \hspace{1em}MCB         & N$_{2:3}$ & N$_{1:2}$ &   --  & 0.24 & 37.037 & -- & -- & -- \\
    \hspace{1em}OCB         & N$_{1:2}$ &  --    &   --  & 0.24 & 37.037 & -- & -- & -- \\
    \hline
  \end{tabular}
  \caption[Definition of SSO populations]{%
    The definition of all the dynamical populations use here, as
    function of their semi-major axis, eccentricity,
    perihelion, and aphelion \citep[using the definitions
      in][]{2016-Icarus-268-Carry, 2008-SSBN-2-Gladman}.
    See Fig.~\ref{fig:pop} for the distribution of these populations in
    the semi-major axis - eccentricity orbital element space.
    The numerical value of the semi-major axes $a$, perihelion $q$,
    aphelion $Q$,
    and mean-motion resonances (Indices $_{i:j}$) are 
    for the Earth 
    a$_\textrm{\earth}$, 
    q$_\textrm{\earth}$, and 
    Q$_\textrm{\earth}$ at 1.0, 0.983, and 1.017 AU; 
    for Mars Q$_\textrm{\mars}$ at 1.666 AU; 
    for Jupiter
    J$_{4:1}$,
    J$_{3:1}$,
    J$_{5:2}$,
    J$_{2:1}$, and 
    J$_{5:3}$ at
    2.06, 2.5, 2.87, 3.27, 3.7 AU;
    and for Neptune
    a$_\textrm{\neptune}$, 
    N$_{2:3}$, and
    N$_{1:2}$ at
    30.07, 47.7, and 39.4 AU.
    The somewhat arbitrary limit of 37.037 AU corresponds to the
    innermost perihelion accessible to detached KBOs (semi-major axis
    of N$_{1:2}$ and eccentricity of 0.24).
    \label{tab:orbit}
  }
\end{table*}

\clearpage
\section{Euclid colors and lightcurves of SSOs\label{app:phot}}
  \indent Due to the ever changing Sun-SSO-observer geometry and 
  SSO rotating irregular shape, the apparent magnitude of SSOs is
  constantly changing. Magnitude variations in multi-filter time
  series are thus a mixture of
  low frequency geometric evolution, 
  high frequency shape-related variability, and
  intrinsic surface colors.

  \indent The slow geometric evolution can easily be taken into
  account (Eq.~\ref{eq:H}), but disentangling the intrinsic surface
  colors from the shape-related
  variability is required to
  build the SED (Section~\ref{sec:taxo}) and to
  obtain a dense lightcurve  (Section~\ref{sec:photo}).
  Often, only the simplistic approach of taking the pair of filters
  closest in time can be used to determine the color
  \citep[e.g.,][]{2016-AA-591-Popescu}, while hoping the shape-related
  variability will not affect the color measurements
  \citep[Fig.~\ref{fig:lc},][]{2004-MNRAS-348-Szabo}. 

  \indent The sequence of observation by Euclid in four repeated
  blocks, each containing all four filters (Fig.~\ref{fig:seq}),
  however allows a more subtle approach.
  For any given color, i.e., pair of filter, each filter will be
  bracketed in time three times by the other filter. The reference
  magnitudes provided by the bracketing filter allow to estimate the
  magnitude at the observing time of the other filter.
  For instance, to determine the (VIS-Y) index, one can use the first
  two measurements in VIS to estimate what should be the VIS magnitude
  at the time the Y filter was acquired (by simple linear
  interpolation for instance). This corrects, although only partially,
  for the shape-related variability.
  Hence, any colors
  will be evaluated six times over an hour, although not entirely
  independently each time. 

  \indent The only notable assumption here is that the SED is constant
  over rotation, i.e., that the surface composition and properties are
  homogeneous on the surface, which is a soft assumption based on the history of 
  spacecraft rendezvous with asteroids 
  \citep[i.e., Eros, Gaspra, Itokawa, Mathilde, Ida, \v{S}teins,
    Lutetia, Ceres, with the only exception of
    the Vesta, see e.g.,][]{
    2000-Science-289-Veverka,
    2011-Science-334-Sierks,
    2012-Science-336-Russell}.

\begin{figure*}[h]
  \centering
  \includegraphics[width=.5\hsize]{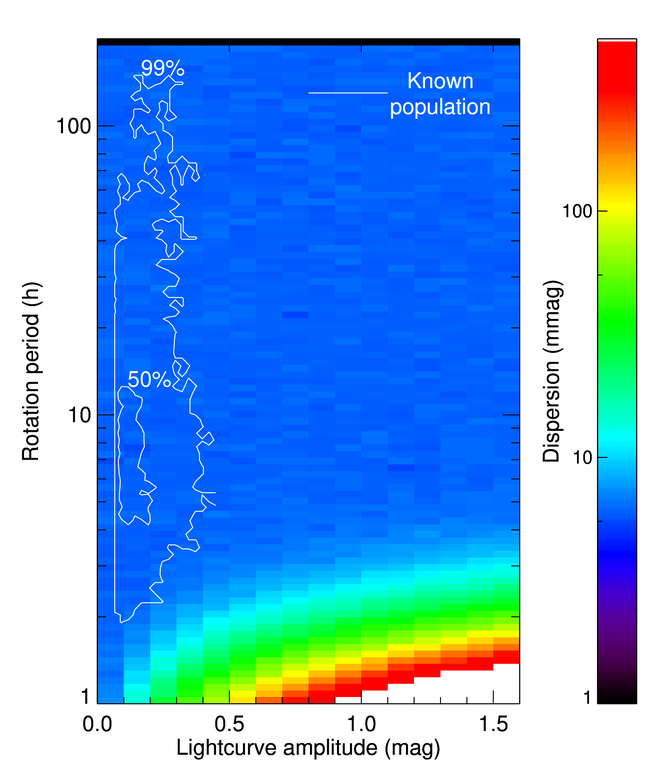}%
  \includegraphics[width=.5\hsize]{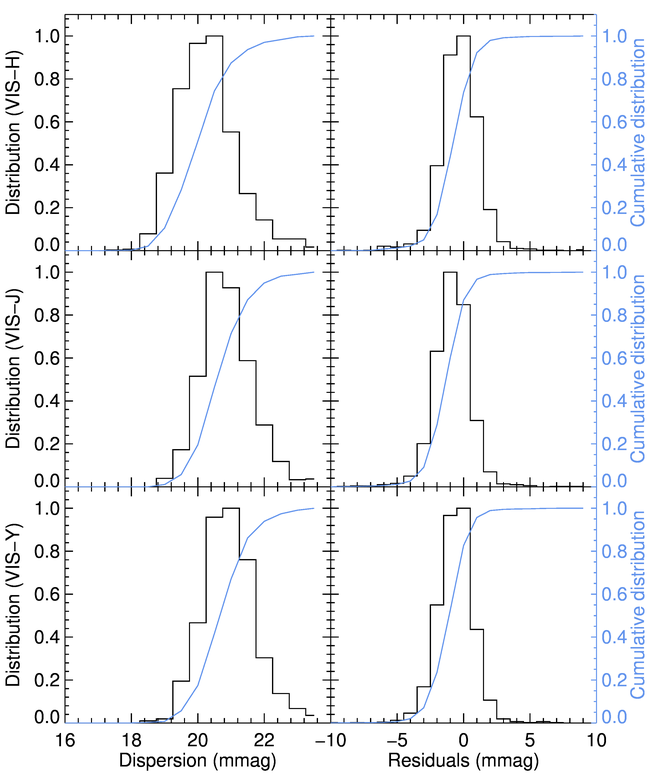}%
  \caption[Colors residuals and dispersion from simulations]{
    \textbf{Left:} Distribution of the dispersion of color measurement
    in period-amplitude space.
    The white contours represent the regions encompassing
    respectively 50\% and 99\% of the population with known
    rotation period and amplitude.
    Largest uncertainties are found for high-amplitude
    short-rotation-period lightcurves, outside the typical space
    sampled by SSOs.
    \textbf{Right:}
    Distribution of the dispersion and residuals of color
    determination in VIS-Y, VIS-J, and VIS-H colors (the remaining colors
    being a combination of these three).
    Dispersion is typically at the level of individual measurement
    uncertainty (here 0.020 magnitude). Residuals are much smaller,
    close to zero and with dispersion below 0.01 magnitude.
  }
  \label{fig:avp}
\end{figure*}

  \indent We test this approach by simulating sequences of observation
  by Euclid. For each of the 371 asteroids of the
  \citet{2009-Icarus-202-DeMeo}, we simulate 800 lightcurves made of
  Fourier series of the second order, with random coefficients to produce
  lightcurve amplitude between 0 and 1.6 magnitude, and random
  rotation period between 1 and 200 hours.
  These $\approx$300,000 lightcurves span the observed range of
  amplitude and period parameter space, estimated from the 5759
  entries with a quality code 2 or 3 from the Planetary Data System archive
  \citep[Fig.~\ref{fig:avp},][]{2017-PDS-Harris}.
  We limit the simulation to second order Fourier series as dense
  lightcurves for about a thousand asteroids from the Palomar
  Transient Factory showed that is was sufficient to reproduce most
  asteroid lightcurves
  \citep{2012-MNRAS-421-Polishook, 2014-ApJ-788-Chang,
    2015-AJ-150-Waszczak}. 
  For each lightcurve, we determine the 4$\times$4 apparent magnitude
  measurements using the definition of Euclid observing sequence
  (Fig.~\ref{fig:seq}), the SSO color (from Section~\ref{sec:taxo}), and
  add a random Gaussian noise of 0.02 magnitude. 

  \indent We then analyze these 4$\times$4 measurements with the
  method described above. For each SSO and each lightcurve, we
  determine all the colors (VIS-Y, VIS-J, VIS-H, Y-J, Y-H, J-H) and
  compare them with the input of the simulation, hereafter the
  \textsl{residuals}.
  For each color, we also record the \textsl{dispersion} of
  estimates. 

  \indent The accuracy on each colors is found to be at the level of
  single measurement uncertainty (Fig.~\ref{fig:avp}). This is due to
  the availability of multiple estimates of each color, improving the
  resulting signal to noise ratio.
  The residuals are found very close to zero: offsets below the
  milli-magnitude (mmag) with a standard deviation below 0.01, i.e.,
  smaller than 
  individual measurement uncertainty (about a factor of five). 
  We repeated the analysis with higher levels of Gaussian noise on
  individual measurements (0.05 and 0.10 magnitude, the latter corresponding to
  the expected precision at Euclid limiting magnitude), adding 600,000
  simulated lightcurves to the exercise, and found similar results:
  color uncertainty remains at the level of the uncertainty on
  individual measurement, and residuals remain close to zero, with a
  dispersion following the individual
  measurement uncertainty reduced by a factor of about five.
  The colors determined with this
  technique are therefore precise and reliable.  

  \indent The processing described here is a simple demonstrator that
  SED can be precisely determined from Euclid multi-filters time
  series. As a corollary, a single lightcurve of 16 measurements can
  be reconstructed from the 4$\times$4 measurements.
  These will be the root of the spectral classification
  (Section~\ref{sec:taxo}) and time-resolved photometry analysis
  (Section~\ref{sec:photo}).
  The technique will be further refined for the data processing: 
  we considered here each color, i.e. pair of filters, independently. No 
  attempt for multi-pair analysis was made for this simple
  demonstration of the technique, while a combined analysis should
  reduce even further the residuals, i.e., potential biases.

\end{document}